\documentclass[12pt]{article}

\usepackage{epsfig}
\usepackage{psfrag}
\usepackage{latexsym}
\usepackage{indentfirst}
\usepackage{fancyhdr}
\usepackage{dsfont}
\usepackage{amssymb}
\usepackage{amsmath}
\usepackage{amsfonts}
\usepackage{pifont}
\usepackage{cite}
\usepackage{bbold}
\usepackage{color}
\usepackage{colordvi}
\usepackage{fancybox}
\usepackage[footnotesize]{caption2}
\usepackage{graphicx}
\usepackage[center,footnotesize,hang]{subfigure}
\usepackage{bbm}
\usepackage{url}
\usepackage{multirow}
\usepackage{array}
\usepackage{arydshln}
\usepackage{ulem}
\newcommand{\PreserveBackslash}[1]{\let\temp=\\#1\let\\=\temp}
\newcolumntype{C}[1]{>{\PreserveBackslash\centering}p{#1}}
\newcolumntype{R}[1]{>{\PreserveBackslash\raggedleft}p{#1}}
\newcolumntype{L}[1]{>{\PreserveBackslash\raggedright}p{#1}}
\newcommand{\cleqn}{\setcounter{equation}{0}}

\newcommand{\bq}{\begin{eqnarray}}
\newcommand{\nq}{\end{eqnarray}}

\begin{document}

\title{\hfill ~\\[0mm]
        \textbf{Generalised CP and Trimaximal TM$_1$ Lepton Mixing in $S_4$ Family Symmetry}}
\date{}

\author{\\[1mm]Cai-Chang Li\footnote{E-mail: {\tt lcc0915@mail.ustc.edu.cn}}~,~Gui-Jun Ding\footnote{E-mail: {\tt dinggj@ustc.edu.cn}}\\ \\
\it{\small Department of Modern Physics, University of Science and
    Technology of China,}\\
  \it{\small Hefei, Anhui 230026, China}\\[4mm] }
\maketitle

\begin{abstract}

We construct two flavor models based on $S_4$ family symmetry and generalised CP symmetry. In both models, the $S_4$ family symmetry is broken down to the $Z^{SU}_2$ subgroup in the neutrino sector, as a consequence, the trimaximal $\text{TM}_1$ lepton mixing is produced. Depending on the free  parameters in the flavon potential, the Dirac CP is predicted to be either conserved or maximally broken, and the Majorana CP phases are trivial. The two models differ in the neutrino sector. The flavon fields are involved in the Dirac mass terms at leading order in the first model, and the neutrino mass matrix contains three real parameters such that the absolute neutrino masses are fixed. Nevertheless, the flavon fields enter into the Majorana mass terms at leading order in the second model. The leading order lepton mixing is of the tri-bimaximal form which is broken down to $\text{TM}_1$ by the next to leading order contributions.

\end{abstract}
\thispagestyle{empty}
\vfill

\newpage
\setcounter{page}{1}

\section{\label{sec:1} Introduction}
\cleqn

In the past years, the Daya Bay~\cite{An:2012eh}, RENO~\cite{Ahn:2012nd} and Double Chooz~\cite{Abe:2011fz} experiments, together with the long-baseline experiments T2K~\cite{Abe:2011sj} and MINOS~\cite{Adamson:2011qu}, have provided an accurate determination of the last unknown lepton mixing angle $\theta_{13}$, with the latest central
value measured by Daya Bay being $\theta_{13}\simeq8.7^{\circ}$~\cite{An:2013zwz}. The measurement of the reactor angle excluded many neutrino mass models, and led to new model building strategies based on family symmetries~\cite{Toorop:2011jn,Ding:2012xx,King:2012in,Ding:2012wh}. So far all the three lepton mixing angles and both mass-squared differences $\Delta m_{sol}^2$ and $\Delta m_{atm}^2$ have been measured to reasonably good accuracy. However, barely nothing is known on the leptonic CP phases, which contain one Dirac phase $\delta_{CP}$ and two Majorana phases $\alpha_{21}$ and $\alpha_{31}$. The global analysis of the current neutrino oscillation data gives that the $3\sigma$ range of $\delta_{CP}$ is $\left[0, 2\pi\right)$~\cite{Tortola:2012te,Fogli:2012ua,GonzalezGarcia:2012sz}, although there is some indications for non-zero $\delta_{CP}$. Therefore we still don't know whether CP violation occurs in the lepton
sector and how large it is if the CP symmetry is really violated. Measuring the leptonic CP violation is one of important goals of future long-baseline neutrino oscillation experiments~\cite{Bass:2013vcg}.

Family symmetry and its spontaneous breaking have turned out to be able to naturally derive some mass independent textures, please see Ref.~\cite{Altarelli:2010gt} for a review. In order to explain the observed lepton mixing angles and predict CP phases at the same time, it is natural to extend the family symmetry to include a generalised CP symmetry~\cite{Ecker:1981wv,Feruglio:2012cw,Holthausen:2012dk,Ding:2013hpa,Feruglio:2013hia,Luhn:2013lkn,Ding:2013bpa}. In this setup, the symmetries are spontaneously broken by the flavon vacuum expectation values (VEVs) which take specific discrete complex phases. As in the paradigm of family symmetry, the whole symmetry including both family and CP symmetries are generally broken into different subgroups in the neutrino and charged lepton sectors, and the mismatch between the two remnant subgroups gives rise to particular predictions for lepton mixing angles and CP phases.

Combining family symmetry with generalised CP symmetry is a promising framework to predict the values of CP violating phases. It has arisen some interesting discussions in the past years. Imposing generalised CP symmetry within the context of simple $\mu-\tau$ interchange symmetry~\cite{mu_tau_reflection}, $A_4$~\cite{Ding:2013bpa}, $S_4$~\cite{Ding:2013hpa,Feruglio:2013hia,Luhn:2013lkn,Krishnan:2012me,Mohapatra:2012tb} and $T^{\prime}$~\cite{Girardi:2013sza} family symmetries have been explored (other approaches to discrete symmetry and CP violation
can be found in Refs.~\cite{Branco:1983tn,Chen:2009gf,Antusch:2011sx}). In such scenario, the mixing angles and CP phases are generally predicted to be strongly correlated with each other because of the constraint of the family and CP symmetries. The so-called trimaximal $\text{TM}_2$ neutrino mixing, whose second column of the mixing matrix is of the form $(1,1,1)^{T}/\sqrt{3}$, is frequently produced. In Refs.~\cite{Ding:2013hpa,Ding:2013bpa}, the $\text{TM}_2$ mixing is a natural consequence of the preserved $Z^{S}_2$
family symmetry in the neutrino sector. In the present work, we shall focus on the trimaximal $\text{TM}_1$ mixing whose first column of the mixing matrix takes the form $(2,-1,-1)^{T}/\sqrt{6}$, since the $\text{TM}_1$ mixing leads to better agreement of solar mixing angle $\theta_{12}$ with the measured value than the $\text{TM}_2$ pattern. We shall construct two typical models based on $S_4$ family symmetry and the corresponding generalised CP symmetry. 
Both models predict $\text{TM}_1$ mixing due to the remnant $Z^{SU}_2$ symmetry in the neutrino sector and the Dirac CP is conserved or maximally broken. In the first model (Model 1), the flavon fields enter into the neutrino Dirac couplings instead of the Majorana mass terms of the right-handed neutrinos at leading order (LO), and the $\text{TM}_1$ mixing is generated at LO. After taking into account the measured solar and atmospheric neutrino mass squared differences and the reactor mixing angle $\theta_{13}$, the absolute neutrino masses and the effective mass $\left|m_{\beta\beta}\right|$ for the neutrinoless double beta decay are fixed completely. For the second model (Model 2), the lepton mixing is of the tri-bimaximal form at LO with the remnant $Z^{S}_2\times Z^{SU}_2$ family symmetry in the neutrino sector, and the next-to-leading order (NLO) corrections further breaks $Z^{S}_2\times Z^{SU}_2$ into $Z^{SU}_2$ such that $\text{TM}_1$ pattern is obtained. Since the non-zero $\theta_{13}$ arises from the NLO contributions, the relative smallness of the reactor angle with respect to the solar and atmospheric mixing angles are explained.

The paper is organized as follows. In section~\ref{sec:model_independent}, we briefly review the concept of generalised CP symmetry and the generalised CP transformation compatible with $S_4$ family symmetry. Moreover, the possible residual CP symmetries consistent with the remnant $Z^{SU}_2$ family symmetry in the neutrino sector and the corresponding phenomenological predictions for the lepton mixing parameters are investigated. In section~\ref{sec:model_dirac}, we present the first model, and we show that the desired vacuum configuration with their phase structure can be realized in a supersymmetric context. In section~\ref{sec:model_majorana}, we specify our second model, the LO structure of the model, the vacuum alignment and the NLO corrections induced by higher dimensional operators are discussed. We summarize and conclude in section~\ref{sec:conclusion}. The details of the group theory of $S_4$ are given in the Appendix, where the explicit representation matrices and the Clebsch-Gordan coefficients are listed.

\section{\label{sec:model_independent}General analysis of lepton mixing with residual $Z^{SU}_2$ family symmetry and CP symmetry}

\cleqn

\subsection{\label{sec:GCP_S4}Generalised CP transformations consistent with $S_4$}

It is highly non-trivial to combine a family symmetry $G_f$ with the generalised CP symmetry together~\cite{Feruglio:2012cw,Holthausen:2012dk,Ding:2013hpa}. Let us consider a generic multiplet of fields $\varphi(x)$ in the irreducible representation $\mathbf{r}$ of $G_f$. Under the action of $G_f$, $\varphi(x)$  transforms as
\begin{equation}
\varphi\stackrel{G_f}{\longrightarrow} \rho_{\mathbf{r}}(g) \varphi(x),
\qquad g \in G_f\,,
\end{equation}
where $\rho_{\mathbf{r}}(g)$ is the representation matrix for the
element $g$ in the irreducible representation $\mathbf{r}$. The generalised CP transformation should leave the kinetic term $\left|\partial\varphi\right|^2$ invariant and it acts on $\varphi(x)$ as
\begin{equation}
\varphi(x)\stackrel{CP}{\longrightarrow}X_{\bf r}\varphi^{*}(x^{\prime})\,,
\end{equation}
where $X_{\bf r}$ is a unitary matrix, $x^\prime=(t,-\bf{x})$ and we have omit the action of CP on spinor indices for the case that $\varphi$ is a spinor. Notice that we are considering the ``minimal'' theory in which the generalised CP transformation maps the field $\varphi\sim\mathbf{r}$ into its complex conjugate $\varphi^{*}\sim\mathbf{r}^{*}$.
The generalised CP transformation $X_{\mathbf{r}}$ has to be consistently defined to be compatible with the family symmetry $G_f$. Hence the so-called consistency condition~\cite{Grimus:1995zi,Feruglio:2012cw,Ding:2013hpa,Holthausen:2012dk} must be satisfied
\begin{equation}
\label{eq:consistency}
X_{\mathbf{r}}\rho_{\mathbf{r}}^{*}(g)X_{\mathbf{r}}^{-1}=\rho_{\mathbf{r}}(g^{\prime}),~~~~ g,g^{\prime}\in G_f\,.
\end{equation}
Note that Eq.~\eqref{eq:consistency} should be fulfilled for all the irreducible representations of $G_f$. Moreover, Eq.~\eqref{eq:consistency} implies that the generalised CP transformation $X_{\mathbf{r}}$ maps the group element $g$ into $g'$, and this mapping preserves the family symmetry group structure~\cite{Holthausen:2012dk,Ding:2013hpa}. Therefore Eq.~\eqref{eq:consistency} defines a homomorphism of the family symmetry group $G_f$. It is now established that there is one to one correspondence
between the generalised CP transformations and the automorphism group of the family symmetry group~\cite{Grimus:1995zi}.

In the present work, we shall concentrate on the family symmetry $G_f=S_4$, which can be generated by three generators $S$, $T$ and $U$. It is convenient to work in the $T$ generator diagonal basis, the representation matrices for the three generators in different $S_4$ irreducible representations are summarized in Table~\ref{tab:representation}. The corresponding Clebsch-Gordan coefficients are listed in Appendix A. The automorphism structure of $S_4$ is rather simple, since it doesn't have non-trivial outer automorphism. Therefore the automorphism of $S_4$ is exactly its inner automorphism, and the automorphism group of $S_4$ is isomorphic to $S_4$ itself. For the representative automorphism element $\text{conj}(U):(S,T,U)\rightarrow(S,T^2,U)$, 
where \text{conj}(h) denotes a group conjugation with an element $h$, i.e. \text{conj}(h): $g\rightarrow hgh^{-1}$ with $h, g\in S_4$,
the associated generalised CP transformation $X^{0}_{\mathbf{r}}$ is determined by the consistency equations
\begin{equation}
\label{eq:relationship-one}
X^{0}_{\mathbf{r}}\rho_{\mathbf{r}}^{*}(S)\left(X_{\mathbf{r}}^{0}\right)^{-1}=\rho_{\mathbf{r}}(S),\quad X^{0}_{\mathbf{r}}\rho_{\mathbf{r}}^{*}(T)\left(X_{\mathbf{r}}^{0}\right)^{-1}=\rho_{\mathbf{r}}(T^{2}),\quad X^{0}_{\mathbf{r}}\rho_{\mathbf{r}}^{*}(U)\left(X_{\mathbf{r}}^{0}\right)^{-1}=\rho_{\mathbf{r}}(U)\,.
\end{equation}
From the explicit form of the representation matrices shown in Table~\ref{tab:representation}, we see that for any irreducible representations $\mathbf{r}$ of $S_4$, the following relations are fulfilled
\begin{equation}
\label{eq:relationship-two}
\rho_{\bf r}^*(S)=\rho_{\bf r}(S),\quad \rho_{\bf r}^*(U)=\rho_{\bf r}(U),\quad \rho_{\bf r}^*(T)=\rho_{\bf r}(T^2)\,.
\end{equation}
Therefore $X^{0}_{\mathbf{r}}$ is fixed to be equal to identity (up to an arbitrary overall phase), i.e.
\begin{equation}
X^{0}_{\mathbf{r}}=1\,.
\end{equation}
Including the family symmetry transformation, the generalised CP transformation consistent with the $S_4$ family symmetry is given by
\begin{equation}
X_{\mathbf{r}}=\rho_{\mathbf{r}}(g)X^{0}_{\mathbf{r}}=\rho_{\mathbf{r}}(g),\quad g\in S_4\,.
\end{equation}
Hence the generalised CP transformation consistent with an $S_4$ family symmetry is of the same form as the family group transformation in the chosen basis. We confirm the results in Refs.~\cite{Holthausen:2012dk,Ding:2013hpa} that the generalised CP transformation group is the identity up to inner automorphism. Since we have found all generalised CP transformations consistent with the $S_4$ family symmetry, we turn to investigate their phenomenological implications on lepton masses and flavor mixings in the following.

\subsection{\label{subsec:general_analysis}Lepton mixing from $S_4\rtimes H_{CP}$ breaking into $G^{l}_{CP}\cong Z^{T}_3\rtimes H^{l}_{CP}$ and $G^{\nu}_{CP}\cong Z^{SU}_2\times H^{\nu}_{CP}$}


In this work we shall introduce the family symmetry $S_4$ together with the corresponding generalised CP symmetry $H_{CP}$ at high energy scale, where $H_{CP}$ is the collection of the generalised CP transformations $X_{\mathbf{r}}$. Hence the original symmetry of the theory is $S_4\rtimes H_{CP}$. To obtain phenomenologically acceptable lepton masses and mixings, the original symmetry should be broken in both charged lepton and neutrino sectors. The mismatch between the symmetry breaking patterns in the neutrino and charged lepton sectors leads to particular predictions for lepton mixing angles and CP phases. In Ref.~\cite{Ding:2013hpa}, the symmetry is broken down to $Z^{S}_2\times H^{\nu}_{CP}$ in the neutrino sector, and the residual family symmetry $Z^{S}_2=\left\{1,S\right\}$ enforces that the lepton mixing is the trimaximal $\text{TM}_2$ pattern~\cite{Albright:2010,Hexiaogang:2011,Rodejohann:2012cf}, where the second column of the Pontecorvo-Maki-Nakagawa-Sakata (PMNS) matrix is proportional to $(1,1,1)^T$. In the present work, we shall investigate another case that $S_4\rtimes H_{CP}$ is broken to $G^{l}_{CP}\cong Z^{T}_3\rtimes H^{l}_{CP}$ and $G^{\nu}_{CP}\cong Z^{SU}_2\times H^{\nu}_{CP}$ in the charged lepton and the neutrino sectors respectively, where $Z^{T}_3=\left\{1, T, T^2\right\}$ and $Z^{SU}=\left\{1, SU\right\}$. The remnant $Z^{SU}_2$ symmetry would lead to the trimaximal $\text{TM}_1$ mixing pattern~\cite{Varzielas:2012pa,Grimus:2013rw,Luhn:2013lkn}, where the first column of the PMNS matrix is proportional to $(2,-1,-1)^T$. General phenomenology analysis has shown that $\text{TM}_1$ mixing can lead to excellent agreement with the present data~\cite{Rodejohann:2012cf}. Furthermore, if the residual family symmetry in the neutrino sector chosen to be $Z^{U}_{2}=\left\{1,U\right\}$, the third column of the mixing matrix would be proportional to $(0,1,-1)^{T}$. The reactor mixing angle would be predicted to be zero, and it is not consistent with both the experimental measurements~\cite{An:2012eh,Ahn:2012nd,Abe:2011fz,Abe:2011sj,Adamson:2011qu,An:2013zwz} and the global data fitting~\cite{Tortola:2012te,Fogli:2012ua,GonzalezGarcia:2012sz}. Hence we don't consider this scenario.

In the charged lepton sector, the full symmetry $S_4\rtimes H_{CP}$ is broken to $G^{l}_{CP}\cong Z^{T}_3\rtimes H^{l}_{CP}$. For $G^{l}_{CP}$ to be a well-defined symmetry, the consistency condition of Eq.~\eqref{eq:consistency} should be satisfied for the residual family symmetry subgroup $Z^{T}_3$, i.e. the element $X_{\mathbf{r}l}$ of $H^{l}_{CP}$ should fulfill
\begin{eqnarray}
X_{\mathbf{r}l}\rho_{\mathbf{r}}^{*}(T)X_{\mathbf{r}l}^{-1}=\rho_{\mathbf{r}}(T^{\prime})\,,\qquad T^{\prime} \in Z_3^T=\{1,T,T^2\}\,.
\end{eqnarray}
It is easy to check that the remnant CP symmetry $H_{CP}^l$ can take the value
\begin{eqnarray}
H_{CP}^l=\left\{\rho_{\mathbf{r}}(1),\rho_{\mathbf{r}}(T),\rho_{\mathbf{r}}(T^2),\rho_{\mathbf{ r}}(U),\rho_{\mathbf{r}}(TU),\rho_{\mathbf{r}}(T^2U)\right\}\,.
\end{eqnarray}
Without loss of generality, we assume that the three generations of the left-handed lepton doublets are unified into the three-dimensional representation $\mathbf{3}$. The same results would be obtained if the lepton doublets were assigned to $\mathbf{3}^{\prime}$ of $S_4$, since the representation $\mathbf{3}^{\prime}$ differs from $\mathbf{3}$ only in the overall sign of the generator $U$. The charged lepton mass matrix $m_l$ is constrained by the remnant family symmetry $Z^{T}_3$ and the remnant CP symmetry $H^{l}_{CP}$ as
\begin{subequations}
\begin{eqnarray}
\label{eq:remnant_flavor_ch}&&\rho^{\dagger}_{\mathbf{3}}(T)m_{l}m^{\dagger}_{l}\rho_{\mathbf{3}}(T)=m_{l}m^{\dagger}_{l}\,,\\
\label{eq:remnant_CP_ch}&&X^{\dagger}_{\mathbf{3}l}m_{l}m^{\dagger}_{l}X_{\mathbf{3}l}=\left(m_{l}m^{\dagger}_{l}\right)^{*}\,,
\end{eqnarray}
\end{subequations}
where the charged lepton mass matrix $m_l$ is given in the convention in which the left-handed (right-handed) fields are on the left-hand (right-hand) side of $m_l$. Since the representation matrix $\rho_{\mathbf{3}}(T)$ is diagonal, the invariant condition Eq.~\eqref{eq:remnant_flavor_ch} under $Z^{T}_3$ implies that $m_{l}m^{\dagger}_{l}$ is diagonal with
\begin{equation}
m_{l}m^{\dagger}_{l}=\text{diag}\left(m^2_{e},m^2_{\mu},m^2_{\tau}\right)\,,
\end{equation}
where $m_e$, $m_{\mu}$ and $m_{\tau}$ denote the electron, muon and tau masses, respectively. For the case of $X_{\mathbf{r}l}=\left\{\rho_{\mathbf{r}}(1), \rho_{\mathbf{r}}(T), \rho_{\mathbf{r}}(T^2)\right\}$, the conditions of Eq.~\eqref{eq:remnant_CP_ch} is satisfied automatically, and therefore no additional constraints are required. For the remaining values $X_{\mathbf{r}l}=\left\{\rho_{\mathbf{r}}(U), \rho_{\mathbf{r}}(TU), \rho_{\mathbf{r}}(T^2U)\right\}$, the residual CP invariant condition of Eq.~\eqref{eq:remnant_CP_ch} implies $m_{\mu}=m_{\tau}$. Hence this case is not viable phenomenologically. We note that in the models constructed in sections \ref{sec:model_dirac} and \ref{sec:model_majorana}, the $Z^{T}_3$ remnant symmetry is broken by the flavon VEVs in order to facilitate the generation of the charged lepton mass hierarchies without fine tuning. However, we properly arrange the breaking such that the resulting charged lepton mass matrix remains diagonal. As a consequence, the hermitian product $m_{l}m^{\dagger}_l$ is invariant under the action of $Z^{T}_3$ elements and the generalised CP transformations $X_{\mathbf{r}l}=\left\{\rho_{\mathbf{r}}(1), \rho_{\mathbf{r}}(T), \rho_{\mathbf{r}}(T^2)\right\}$, i.e. Eqs.~(\ref{eq:remnant_flavor_ch},\ref{eq:remnant_CP_ch}) are satisfied. Therefore the following general analysis is still meaningful and valid, and in particular it guides our model building.

Now we turn to the neutrino sector. In order to reproduce the $\text{TM}_1$ mixing pattern, the symmetry $S_4\rtimes H_{CP}$ is spontaneously broken to $G_{CP}^\nu=Z_{2}^{SU}\times H_{CP}^{\nu}$. The residual CP symmetry $H_{CP}^{\nu}$ should be consistent with the residual family symmetry $Z_{2}^{SU}$, and therefore its element $X_{\mathbf{r}\nu}$ has to fulfill the consistency equation
\begin{eqnarray}
\label{eq:consistent-two}
X_{\mathbf{r}\nu}\rho_{\mathbf{r}}^{*}(SU)X_{\mathbf{r\nu}}^{-1}=\rho_{\mathbf{r}}(SU)\,.
\end{eqnarray}
One can easily check that there are only 4 possible choices for $X_{\mathbf{r\nu}}$, i.e.
\begin{equation}
H^{\nu}_{CP}=\left\{\rho_{\mathbf{r}}(1), \rho_{\mathbf{r}}(S), \rho_{\mathbf{r}}(U), \rho_{\mathbf{r}}(SU)\right\}\,.
\end{equation}
The light neutrino mass matrix $m_{\nu}$ is constrained by the residual family symmetry $Z^{SU}_2$ and residual CP symmetry $H_{CP}^{\nu}$ as~\cite{Ding:2013hpa}
\begin{subequations}
\begin{eqnarray}
\label{eq:remnant_flavor_nu} && \rho_{\mathbf{3}}^{T}(SU)m_{\nu}\rho_{\mathbf{3}}(SU)=m_{\nu}\,, \\
\label{eq:remnant_CP_nu} && X_{\mathbf{3}\nu}^Tm_{\nu}X_{\mathbf{3}\nu}=m^{*}_{\nu}\,.
\end{eqnarray}
\end{subequations}
The most general neutrino mass matrix which satisfies Eq.~\eqref{eq:remnant_flavor_nu} is of the form
\begin{eqnarray}
\label{eq:general-mass-matrix}
m_{\nu}=\alpha\left(\begin{array}{ccc}
2  &  -1  &  -1  \\
-1  &  2  &  -1  \\
-1  &  -1  &  2
\end{array}\right)+\beta\left(\begin{array}{ccc}
1  &  0  &  0  \\
0  &  0  &  1  \\
0  &  1  &  0
\end{array}\right)+\gamma\left(\begin{array}{ccc}
0  &  1  &  1  \\
1  &  1  &  0  \\
1  &  0  &  1
\end{array}\right)+\delta\left(\begin{array}{ccc}
0  &  1  &  -1  \\
1  &  2  &  0  \\
-1  & 0   &  -2
\end{array}\right)\,,
\end{eqnarray}
where the four parameters $\alpha$, $\beta$, $\gamma$ and $\delta$ are generally complex, and the remnant CP invariant condition of Eq.~\eqref{eq:remnant_CP_nu} would further constrain these parameters to be real or purely imaginary.

In order to diagonalize light neutrino mass matrix $m_{\nu}$ in Eq.~\eqref{eq:general-mass-matrix}, it is useful to first perform a tri-bimaximal transformation $U_{TB}$
\begin{eqnarray}\label{eq:general-mass-matrix-two}
m_{\nu}^{\prime}=U_{TB}^{T}m_{\nu}U_{TB}= \left(
\begin{array}{ccc}
 3 \alpha+\beta-\gamma  & 0 & 0 \\
 0 & \beta+2\gamma  & -\sqrt{6}\;\delta  \\
 0 & -\sqrt{6}\;\delta  & ~3\alpha-\beta+\gamma
\end{array}
\right)\,,
\end{eqnarray}
with
\begin{equation}
U_{TB}=\left(\begin{array}{ccc}
\sqrt{\frac{2}{3}}  &   \frac{1}{\sqrt{3}}  &  0  \\
-\frac{1}{\sqrt{6}}  &  \frac{1}{\sqrt{3}}  &  -\frac{1}{\sqrt{2}}  \\
-\frac{1}{\sqrt{6}}  &  \frac{1}{\sqrt{3}}  &  \frac{1}{\sqrt{2}}
\end{array}\right)\,.
\end{equation}
Then we investigate the implication of the remnant CP invariant condition of Eq.~\eqref{eq:remnant_CP_nu}. Two distinct phenomenological predictions arise for $X_{\mathbf{r}\nu}=\{\rho_{\mathbf{r}}(1),\rho_{\mathbf{r}}(SU)\}$ and $X_{\mathbf{r}\nu}=\{\rho_{\mathbf{r}}(1),\rho_{\mathbf{r}}(SU)\}$. We shall discuss the two cases in detail in the following.
\begin{description}
\item[~~(\uppercase\expandafter{\romannumeral1})]
$X_{\mathbf{r}\nu}=\rho_{\mathbf{r}}(1), \rho_{\mathbf{r}}(SU)$ 

In this case, we can straightforwardly find that all the four parameters $\alpha$, $\beta$, $\gamma$ and $\delta$ are constrained to be real. As a result, $m_{\nu}^{\prime}$ becomes a real symmetry matrix and can be diagonalized by a rotation matrix $R(\theta)$ in the $(2,3)$ sector with
\begin{equation}\label{eq:orthogonal}
R(\theta)=\left(
\begin{array}{ccc}
 1 & 0 & 0 \\
 0 &  \cos ~\theta  & \sin \theta \\
 0 & -\sin \theta & \cos ~\theta
\end{array}
\right)\,,
\end{equation}
where
\begin{equation}\label{eq:rotation-one}
\tan2\theta=\frac{-2\sqrt{6}\;\delta}{3\alpha-2\beta-\gamma}\,.
\end{equation}
Hence we have
\begin{equation}
\label{eq:diagonal-two}
U_{\nu}^{\prime T}m_{\nu}^{\prime}U_{\nu}^{\prime}=\text{diag}(m_1,m_2,m_3),\qquad U^{\prime}_{\nu}=R(\theta)P\,,
\end{equation}
where $P$ is a unitary diagonal matrix with entries $\pm1$ or $\pm i$, which encode the CP parity of the neutrino state. Furthermore, the light neutrino masses $m_{1,2,3}$ are determined to be
\begin{eqnarray}\label{eq:mass-one}
\nonumber &&\hskip-0.2in m_1=\left|3\alpha+\beta-\gamma\right|\,, \\
\nonumber &&\hskip-0.2in m_2=\frac{1}{2}\left|3(\alpha+\gamma)-\text{sign}\left((3\alpha-2\beta-\gamma)\cos2\theta\right)\sqrt{24 \delta^2+(3\alpha-2\beta-\gamma)^2}\right|\,, \\
&&\hskip-0.2in m_3=\frac{1}{2}\left|3(\alpha+\gamma)+\text{sign}\left((3\alpha-2\beta-\gamma)\cos2\theta\right)\sqrt{24 \delta^2+(3\alpha-2\beta-\gamma)^2}\right|\,.
\end{eqnarray}
We see that the three neutrino masses depend on four real parameters, and therefore any neutrino mass spectrum can be realized in this scenario. Since the charged lepton mass matrix is diagonal, the lepton mixing matrix $U_{PMNS}$ is fixed by neutrino sector completely, and we have
\begin{equation}\label{eq:UPMNS1}
U_{PMNS}=U_{TB}U^{\prime}_{\nu}=\left(
\begin{array}{ccc}
\sqrt{\frac{2}{3}} & \frac{\cos\theta }{\sqrt{3}} & \frac{\sin\theta }{\sqrt{3}} \\
-\frac{1}{\sqrt{6}} & \frac{\cos\theta }{\sqrt{3}}+\frac{\sin\theta }{\sqrt{2}} & -\frac{\cos\theta }{\sqrt{2}}+\frac{\sin\theta }{\sqrt{3}} \\
-\frac{1}{\sqrt{6}} & \frac{\cos\theta }{\sqrt{3}}-\frac{\sin\theta }{\sqrt{2}} & \frac{\cos\theta }{\sqrt{2}}+\frac{\sin\theta }{\sqrt{3}}
\end{array}
\right)P\,.
\end{equation}
The three lepton mixing angles $\theta_{13}$, $\theta_{12}$ and $\theta_{23}$ are predicted to be
\begin{eqnarray}
\label{eq:mixing-angles-one}
\nonumber &&\sin^2\theta_{13}=\frac{1}{3}\sin^2\theta,\quad \sin^2\theta_{12}=\frac{\cos^2\theta}{2+\cos^2\theta}=\frac{1}{3}-\frac{2}{3}\tan ^2\theta _{13}\,,  \\
&&\sin^2\theta_{23}=\frac{1}{2}-\frac{\sqrt{6}\;\sin\theta\cos\theta}{3-\sin^2\theta}=\frac{1}{2}\pm \tan\theta _{13}\sqrt{2(1-2\tan^2\theta _{13})}\,.
\end{eqnarray}
For the best fitting value of the reactor angle $\theta_{13}=8.71^{\circ}$~\cite{GonzalezGarcia:2012sz}, the remaining two mixing angles are determined to be $\theta_{12}\simeq34.31^{\circ}$ and $\theta_{23}\simeq32.49^{\circ}$ or $\theta_{23}\simeq57.51^{\circ}$, which are compatible with the preferred values from global fits. Adopting the PDG parameterization\cite{pdg}, the Dirac CP violating phase $\delta_{CP}$ and two Majorana CP violating phases $\alpha_{21}$ and $\alpha _{31}$ take the values
\begin{equation}
\sin\delta_{CP}=\sin\alpha_{21}=\sin\alpha_{31}=0\,,
\end{equation}
which implies
\begin{eqnarray}
\label{eq:CPphase1}\delta_{CP} ,\alpha_{21} ,\alpha_{31} =0,  ~\pi\,.
\end{eqnarray}
Hence there is no CP violation in this case.
\item[~~(\uppercase\expandafter{\romannumeral2})]   $X_{\mathbf{r}\nu}=\rho_{\mathbf{r}}(S),\rho_{\mathbf{r}\nu}(U)$ 

Solving the residual CP invariant equation of Eq.~\eqref{eq:remnant_CP_nu}, we find the three parameters $\alpha$, $\beta$ and $\gamma$ are real, and $\delta$ is purely imaginary. The unitary transformation $U^{\prime}_{\nu}$ diagonalizing the neutrino mass matrix $m^{\prime}_{\nu}$ is of the form
\begin{equation}
U^{\prime}_{\nu}=\left(\begin{array}{ccc}
1  &    0   &    0  \\
0   &   \cos\theta   &   \sin\theta  \\
0   &  -i\sin\theta  &   ~i\cos\theta
\end{array}\right)P\,,
\end{equation}
with
\begin{equation}
\label{eq:rotation-two}
\tan2\theta=\frac{2i\sqrt{6}\;\delta}{3\left(\alpha+\gamma\right)}\,.
\end{equation}
The resulting PMNS matrix is
\begin{equation}
\label{eq:UPMNS2}
U_{PMNS}=U_{TB}U^{\prime}_{\nu}=\left(
\begin{array}{ccc}
\sqrt{\frac{2}{3}} & \frac{\cos\theta }{\sqrt{3}} & \frac{\sin\theta }{\sqrt{3}} \\
-\frac{1}{\sqrt{6}} & \frac{\cos\theta }{\sqrt{3}}+\frac{i \sin\theta }{\sqrt{2}} & -\frac{i \cos\theta }{\sqrt{2}}+\frac{\sin\theta }{\sqrt{3}} \\
-\frac{1}{\sqrt{6}} & \frac{\cos\theta }{\sqrt{3}}-\frac{i \sin\theta }{\sqrt{2}} & \frac{i \cos\theta }{\sqrt{2}}+\frac{\sin\theta }{\sqrt{3}}
\end{array}
\right)P\,.
\end{equation}
The lepton mixing angles and CP phases are determined to be
\begin{eqnarray}
\label{eq:mixing-angles-two}
\nonumber& \left|\sin\delta_{CP}\right|=1,\quad \sin\alpha_{21}=\sin\alpha_{31}=0 \\
& \sin^2\theta_{13}=\frac{1}{3}\sin^2\theta,\quad \sin^2\theta_{12}=\frac{\cos^2\theta}{2+\cos^2\theta}=\frac{1}{3}-\frac{2}{3}\tan ^2\theta _{13}, \quad \sin^2\theta_{23}=\frac{1}{2}\,.
\end{eqnarray}
The predictions for both the solar and reactor mixing angles are the same as the ones in case (I), and the atmospheric mixing angle is maximal. Moreover, we have maximal Dirac CP violation $\delta_{CP}=\pm\frac{\pi}{2}$, and Majorana phases are trivial with $\alpha_{21},\alpha_{31}=0,\pi$. Finally, the light neutrino masses are given by
\begin{eqnarray}
\label{eq:mass-two}
\nonumber && m_1=|3\alpha+\beta-\gamma|\,,   \\
\nonumber && m_2=\frac{1}{2}\left|-3\alpha+2\beta+\gamma+\text{sign}\left((\alpha +\gamma)\cos2\theta\right)\sqrt{9(\alpha+\gamma)^2-24\delta^2}\right|\,,   \\
&& m_3=\frac{1}{2}\left|-3\alpha+2\beta+\gamma-\text{sign}\left((\alpha+\gamma) \cos2\theta\right)\sqrt{9(\alpha+\gamma)^2-24\delta^2}\right|\,.
\end{eqnarray}
\end{description}
Notice that the above results are exactly the same as those of Ref.~\cite{Feruglio:2012cw}, although we use a different basis in which the generator $T$ is diagonal. The chosen basis in the present paper is particularly suitable to build $\text{TM}_1$ model, since the charged lepton mass matrix is diagonal in this basis and the lepton mixing completely comes from the neutrino sector. Now that we have finished the general analysis, we proceed to construct models to realize these model independent results. Two typical models would be proposed in the following sections. In the first model, the lepton mixing is the $\text{TM}_1$ pattern at LO. In the second model, tri-bimaximal mixing is produced at LO, and it is broken to $\text{TM}_1$ mixing by the NLO corrections. As a consequence, the relative smallness of $\theta_{13}$ with respect to $\theta_{12}$ and $\theta_{23}$ is explained.

\section{\label{sec:model_dirac}Model 1}

\cleqn

\begin{table}[t!]
\centering
\begin{tabular}{|c||c|c|c|c|c|c||c|c|c|c|c||c|c|c|c|}
\hline \hline
Field &   $l$    &  $\nu^c$     &  $e^{c}$     &   $\mu^c$    &    $\tau^c$  &  $h_{u,d}$   &  $\varphi_T$ & $\phi $  &   $\varphi _S$  &   $\eta $   &  $\xi  $    &  $\varphi^{0}_T$  &  $\zeta^0$  &  $\varphi^{0}_S$   &  $\eta ^{0}$  \\ \hline

$S_4$  &   $\mathbf{3}^\prime$  &  $\mathbf{3}^\prime$  &  $\mathbf{1}$  & $\mathbf{1}$   &  $\mathbf{1}$   & $\mathbf{1}$   &   $\mathbf{3}^\prime$ & $\mathbf{2}$ &   $\mathbf{3}$ &  $\mathbf{2}$  &  $\mathbf{1}$    &  $\mathbf{3}'$  &  $\mathbf{1}$ &  $\mathbf{3}$   &   $\mathbf{2}$  \\ \hline

$Z_7$  &   $ \omega_7$   &   $1$    &    $\omega^3_7$    &   $\omega^4_7$   &    $\omega^5_7$  &  1   &  $\omega_7$ & $\omega_7$  &  $\omega^6_7$   &   $\omega^6_7$   & $\omega^6_7$   &  $\omega^5_7$  &  $\omega^5_7$  &  $\omega^2_7$   &  $\omega^2_7$ \\ \hline

$U(1)_R$  &  1  &   1  &   1  &   1  &   1  &  0 &    0  &  0 & 0  & 0  &   0   &  2 &   2 &   2 &   2  \\ \hline \hline
\end{tabular}
\caption{\label{tab:effective-two}The particle contents and their transformation property under the family symmetry $S_4\times Z_7$ and $U(1)_R$, where $\omega_7=e^{\frac{2\pi i}{7}}$.}
\end{table}

In this section, we shall present the first $\text{TM}_1$ model (Model 1) based on $S_4\rtimes H_{CP}$ with the extra symmetry $Z_7\times U(1)_R$.
We shall formulate the model in the framework of type I see-saw mechanism
and supersymmetry (SUSY). Both the three generations of left-handed lepton doublets $l$ and the right-handed neutrinos $\nu^c$ are assigned to transform as $S_4$ triplet $\mathbf{3}^{\prime}$, while the RH charged leptons $e^{c}$, $\mu^c$ and $\tau^c$ are all invariant under $S_4$. The involved fields and their transformation rules under the family symmetry $S_4\times Z_7\times U(1)_R$ are summarized in Table~\ref{tab:effective-two}. Notice that the auxiliary $Z_7$ symmetry separates the flavon fields entering the charged lepton sector at LO from those entering the neutrino sector, and it is also helpful to achieve the charged lepton mass hierarchies and suppress the NLO corrections. Compared with most flavor models in which the flavon fields generally couple to the right-handed neutrinos at LO, the flavons are involved in the neutrino Dirac mass term instead of the Majorana mass term of right-handed neutrino in the present model.

\subsection{\label{sec:4.1}Vacuum alignment}

We adopt the now-standard $F-$term alignment mechanism to arrange the vacuum~\cite{Altarelli:2006}. A continuous $U(1)_{R}$ symmetry related to $R-$parity is generally introduced under which the matter fields carry a $+1$ $R-$charge while the electroweak Higgs and flavon fields are uncharged. In addition, one needs the so-called driving fields carrying two unit of $R-$charge, and hence each term in the superpotential can contain at most one driving field. In the SUSY limit, the minimization of the flavon potential can be achieved simply by ensuring that the $F-$terms of
the driving fields vanish at the minimum. The required driving fields and their transformation rules are listed in Table~\ref{tab:effective-two}.
The LO driving superpotential $w_{d}$ invariant under the family symmetry $S_4\times Z_7$ can be written as
\begin{eqnarray}
\label{eq:wd_model1}w_{d}=w_{d}^{l}+w_{d}^{\nu}\,,
\end{eqnarray}
where $w_d^l$ is the flavon superpotential which contains the flavons only entering into the charged lepton at LO, i.e.
\begin{eqnarray} \label{eq:driving-potential-one}
w_d^l=f_1\left(\varphi_T^0\left(\varphi_T\varphi_T\right)_{\mathbf{3}^{\prime}}\right)_{\mathbf{1}}+f_2\left(\varphi_T^0\left(\phi \varphi_T\right)_{\mathbf{3}^{\prime}}\right)_{\mathbf{1}}+f_3\zeta^0\left(\varphi_T\varphi_T\right)_{\mathbf{1}}+f_4\zeta^0\left(\phi\phi\right)_{\mathbf{1}}\,.
\end{eqnarray}
$w_{d}^{\nu}$ is the superpotential associated with the flavons in the neutrino sector
\begin{eqnarray}
\label{eq:driving-potential-five}
\nonumber w_{d}^{\nu}=&&g_1\left(\varphi_S^0\left(\varphi_S\varphi_S\right)_{\mathbf{3}}\right)_{\mathbf{1}}
+g_2\left(\varphi_S^0\left(\eta\varphi_S\right)_{\mathbf{3}}\right)_{\mathbf{1}}+g_3\xi \left(\varphi_S^0\varphi_S\right)_{\mathbf{1}}+g_4\left(\eta^0\left(\varphi_S\varphi_S\right)_{\mathbf{2}}\right)_{\mathbf{1}}\\
&&+g_5\left(\eta^0\left(\eta\eta\right)_{\mathbf{2}}\right)_{\mathbf{1}}+g_6\xi\left(\eta^0\eta\right)_{\mathbf{1}}\,,
\end{eqnarray}
where $\left(\ldots\right)_{\mathbf{r}}$ denotes the contraction into $S_4$ irreducible representation $\mathbf{r}$ according to the Clebsch-Gordan coefficients presented in Appendix~\ref{sec:A}. We note that the first term vanishes automatically due to the anti-symmetric property of the contraction $\left(\varphi_S\varphi_S\right)_{\mathbf{3}}$. Since we require the theory to be invariant under the generalised CP transformation, then all the couplings $f_i$ and $g_i$ in Eqs.~(\ref{eq:driving-potential-one},\ref{eq:driving-potential-five}) are constrained to be real. We start from the charged lepton sector, and the $F-$term conditions obtained from the driving fields $\varphi^{0}_{T}$ and $\zeta^{0}$ read
\begin{eqnarray}
\nonumber\frac{\partial w_d}{\partial\varphi^{0}_{T_1}
}&=&2f_1\left(\varphi^2_{T_1}-\varphi_{T_2}\varphi_{T_3}\right)+f_2\left(\phi _1\varphi_{T_2}+\phi _2\varphi_{T_3}\right)=0\,,\\
\nonumber\frac{\partial w_d}{\partial\varphi^{0}_{T_2}
}&=&2f_1\left(\varphi^2_{T_2}-\varphi_{T_1}\varphi_{T_3}\right)+f_2\left(\phi _1\varphi_{T_1}+\phi _2\varphi_{T_2}\right)=0\,,\\
\nonumber\frac{\partial w_d}{\partial\varphi^{0}_{T_3}
}&=&2f_1\left(\varphi^2_{T_3}-\varphi_{T_1}\varphi_{T_2}\right)+f_2\left(\phi _1\varphi_{T_3}+\phi _2\varphi_{T_1}\right)=0\,, \\
\frac{\partial w_d}{\partial\zeta^{0}}
&=&f_3(\varphi^2_{T_1}+2\varphi_{T_2}\varphi_{T_3})+2f_4\phi _1\phi _2=0 \,.
\end{eqnarray}
We find two possible solutions for the vacuum (up to $S_4$ transformations). The first one is given by
\begin{eqnarray}
\langle\varphi_T\rangle=\left(\begin{array}{c}
 1  \\
 1  \\
 1
\end{array}\right)v_{T},~~
\langle\phi\rangle=\left(\begin{array}{c}
 1  \\
 -1
\end{array}\right)v_{\phi},~~\text{with}~~v_{T}^2=\frac{2f_4}{3f_3}v_{\phi}^2\,.
\end{eqnarray}
The second solution is
\begin{equation}\label{eq:charged-vacuum-one}
\langle\varphi_T\rangle=\left(\begin{array}{c}
0\\
1\\
0
\end{array}\right)v_T,\qquad \langle\phi \rangle=\left(\begin{array}{c}
0\\
1
\end{array}\right)v_{\phi },\quad \text{with} ~~~v_T=-\frac{f_2}{2f_1}v_{\phi }\ .
\end{equation}
We shall choose the second solution in this work. We note that the phase of $v_{\phi}$ can be absorbed into lepton fields, and therefore we can take both $v_{\phi}$ and $v_T$  to be real, since $f_1$ and $f_2$ are real. Furthermore, $v_{\phi}$ and $v_{T}$ are expected to be of the same order of magnitude without fine tuning among the parameters $f_1$ and $f_2$. 
The vacuum expectation values of $\varphi_S$, $\eta$ and $\xi$, which give  rise to $\text{TM}_1$ mixing in the neutrino sector, are determined by the $F-$terms of the associated driving fields as follows:
\begin{eqnarray}
\nonumber&&\frac{\partial w_d^{\nu}}{\partial \varphi^0_{S_1}}= g_2\left(\eta_1\varphi_{S_{2}}+\eta_2\varphi_{S_{3}}\right)+g_3\xi\varphi_{S_{1}}=0\,,\\
\nonumber&&\frac{\partial w_d^{\nu}}{\partial\varphi^0_{S_2}}=g_2\left(\eta_1\varphi_{S_{1}}+\eta_2\varphi_{S_{2}}\right)+g_3\xi\varphi_{S_{3}}=0\,,\\
\nonumber&&\frac{\partial w_d^{\nu}}{\partial\varphi^0_{S_3}}=g_2\left(\eta_1\varphi_{S_{3}}+\eta_2\varphi_{S_{1}}\right)+g_3\xi\varphi_{S_{2}}=0\,,\\
\nonumber&&\frac{\partial w_d^{\nu}}{\partial\eta^0_1}=g_4\left(\varphi_{S_{3}}^2+2\varphi_{S_{1}}\varphi_{S_{2}}\right)+g_5\eta^2_1+g_6\xi\eta_2=0\,,\\
&&\frac{\partial w_d^{\nu}}{\partial\eta^0_2}=g_4\left(\varphi_{S_{2}}^2+2\varphi_{S_{1}}\varphi_{S_{3}}\right)+g_5\eta^2_2+g_6\xi\eta_1=0\,.
\end{eqnarray}
There are two independent solutions to this set of equations up to $S_4$ family symmetry transformations. The first solution is
\begin{equation}
\label{eq:vacuum_neutrino_1st}\langle\varphi_S\rangle=\left(\begin{array}{c}
 1 \\
 1\\
 1
\end{array}\right)v_S,\qquad
\langle\eta\rangle=\left(\begin{array}{c}
  1 \\
  1
\end{array}\right)v_{\eta},\qquad
\langle\xi\rangle=v_{\xi}\,.
\end{equation}
The VEVs $v_S$, $v_{\eta}$ and $v_{\xi}$ are related with each other via
\begin{equation}
v^2_{S}=\frac{g_3(2g_2g_6-g_3g_5)}{12g_4g_2^2}v^2_{\xi},\qquad v_{\eta}=-\frac{g_3}{2g_2}v_{\xi}\,,
\end{equation}
where $v_{\xi}$ is undetermined and generally complex. With the representation matrix given in the Appendix, we can straightforwardly check that the $S_4$ family symmetry is broken down to $Z^{S}_2\times Z^{SU}_2$ subgroup by the vacuum alignment of Eq.~\eqref{eq:vacuum_neutrino_1st}. The second solution reads as
\begin{equation}
\label{eq:neutrino-vacuum-two}
\langle \varphi_S\rangle=\left(\begin{array}{c}
   2  \\
  -1 \\
  -1
\end{array}\right)v_S,\qquad
\langle\eta\rangle=\left(\begin{array}{c}
    1 \\
    1
\end{array}\right)v_{\eta},\qquad
\langle\xi\rangle=v_{\xi}\,,
\end{equation}
with
\begin{equation}
\label{eq:VEV_relation}v^2_S=\frac{g_3(g_3g_5+g_2g_6)}{3g_4g_2^2}v^2_{\xi},\qquad v_{\eta}=\frac{g_3}{g_2}v_{\xi}\,.
\end{equation}
We find the $S_4$ family symmetry is spontaneously broken down to $Z^{SU}_2$ subgroup in this case. In order to reproduce the $\text{TM}_1$ pattern, we choose the second solution in the following.
Since all the couplings $g_i$ are real due to the generalised CP invariance, Eq.~\eqref{eq:VEV_relation} implies that the VEVs $v_{\eta}$ and $v_{\xi}$ have the same phase up to $\pi$, and the phase difference between $v_S$ and $v_\xi$ is $0$, $\pi$ or $\pm\frac{\pi}{2}$ depending on the sign of $g_3g_4\left(g_3g_5+g_2g_6\right)$. In addition, it is natural to expect that the three VEVs $v_{\xi}$, $v_{\eta}$ and $v_{S}$ are of the same order of magnitudes. As shall be shown below, the phase of $v_{\xi}$ turns out to be an overall phase of the light neutrino mass matrix, and hence it can be absorbed into the neutrino fields. That is to say we can take $v_{\xi}$ to be real without loss of generality.  As a consequence, the VEV $v_{\eta}$ would be real as well and the VEV $v_S$ is real for the product $g_3g_4(g_3g_5+g_2g_6)>0$ or purely imaginary for $g_3g_4(g_3g_5+g_2g_6)<0$.

Regarding the order of magnitude of the different VEVs, as we shall find in the following, the charged lepton mass hierarchies can be naturally reproduced if $v_{\phi}/\Lambda$ and $v_{T}/\Lambda$ are of order $\lambda^2$, i.e.
\begin{equation}
\frac{v_{\phi}}{\Lambda}\sim\frac{v_{T}}{\Lambda}\sim\lambda^2\,,
\end{equation}
where $\lambda\simeq 0.23$ is the Cabibbo angle. In order to guarantee the stability of the successful LO results under the inclusion of higher dimensional terms, we choose all the VEVs in the model are of the same order $\lambda^2\Lambda$, i.e.
\begin{equation}
\frac{v_S}{\Lambda}\sim\frac{v_{\eta}}{\Lambda}\sim\frac{v_{\xi}}{\Lambda}\sim\lambda^2\,.
\end{equation}
This assumption is frequently used in the family symmetry model building.

\subsection{\label{sec:4.2}The lepton masses and mixing}

The most general superpotential for the charged lepton masses, which is invariant under the family symmetry, is of the form
\begin{eqnarray}
\label{eq:linear-one}
\nonumber  w_l&&=\frac{y_{\tau}}{\Lambda}\left(l\varphi_T\right)_{\mathbf{1}}\tau^{c}h_d+\frac{y_{\mu_1}}{\Lambda^2} \left(l\left(\varphi_T\varphi_T\right)_{\mathbf{3}^{\prime}}\right)_{\mathbf{1}}\mu^{c}h_d
+\frac{y_{\mu_2}}{\Lambda^2}\left(l\left(\phi\varphi_T\right)_{\mathbf{3}^{\prime}}\right)_{\mathbf{1}}\mu^{c}h_d\\
\nonumber&&+\frac{y_{e_1}}{\Lambda^3}\left(l\varphi_T\right)_{\mathbf{1}}\left(\varphi_T\varphi_T\right)_{\mathbf{1}}e^{c}h_d
+\frac{y_{e_2}}{\Lambda^3}\left(\left(l\varphi_T\right)_{\mathbf{2}}\left(\varphi_T\varphi_T\right)_{\mathbf{2}}\right)_{\mathbf{1}}e^ch_d
+\frac{y_{e3}}{\Lambda^3}\left(\left(l\varphi_T\right)_{\mathbf{3}^{\prime}}\left(\varphi_T\varphi_T\right)_{\mathbf{3}^{\prime}}\right)_{\mathbf{1}}e^ch_d\\
\nonumber&&
+\frac{y_{e_4}}{\Lambda^3}\left(\left(l\varphi_T\right)_{\mathbf{3}}\left(\varphi_T\varphi_T\right)_{\mathbf{3}}\right)_{\mathbf{1}}e^ch_d
+\frac{y_{e_5}}{\Lambda^3}\left(\left(l\phi\right)_{\mathbf{3}^{\prime}}\left(\varphi_T\varphi_T\right)_{\mathbf{3}^{\prime}}\right)_{\mathbf{1}}e^ch_d  +\frac{y_{e_6}}{\Lambda^3}\left(\left(l\phi\right)_{\mathbf{3}}\left(\varphi_T\varphi_T\right)_{\mathbf{3}}\right)_{\mathbf{1}}e^ch_d\\
&&+\frac{y_{e_7}}{\Lambda^3}\left(\left(l\varphi_T\right)_{\mathbf{2}}\left(\phi\phi\right)_{\mathbf{2}}\right)_{\mathbf{1}}e^ch_d +\frac{y_{e_8}}{\Lambda^3}\left(l\varphi_T\right)_{\mathbf{1}}\left(\phi\phi\right)_{\mathbf{1}}e^ch_d+\ldots\,,
\end{eqnarray}
where dots represent the higher dimensional operators which will be commented later. Generalised CP symmetry enforces the Yukawa couplings to be real. Due to the constraint of the $Z_7$ symmetry, the electron, muon and tau mass terms are suppressed by $1/\Lambda$, $1/\Lambda^2$ and $1/\Lambda^3$ respectively. With the vacuum alignment of Eq.~\eqref{eq:charged-vacuum-one}, we find the resulting charged lepton mass matrix is diagonal with
\begin{eqnarray}
\nonumber&m_{e}=\left(y_{e_2}-2y_{e_3}+2y_{e_5}\frac{v_\phi}{v_T}+y_{e_7}\frac{v_\phi^2}{v_T^2}\right)\frac{v_T^3}{\Lambda^3}v_{d}\,,\\
&m_{\mu}=\left(2y_{\mu_1}+y_{\mu_2}\;\frac{v_{\phi}}{v_T}\right)\frac{v_T^2}{\Lambda^2}v_d,\qquad m_{\tau}=y_\tau\frac{v_T}{\Lambda}v_d\,,
\end{eqnarray}
in which $v_d=\langle h_d\rangle$ is the VEV of the electroweak Higgs field $h_d$. We see that the observed mass hierarchies among the charged leptons can be generated for $v_{\phi}/\Lambda\sim v_{T}/\Lambda\sim\lambda^2$.
For the vacuum of $\varphi_T$ and $\phi$ in Eq.~\eqref{eq:charged-vacuum-one}, we can check that the $S_4$ family symmetry is broken completely in the charged lepton sector, since $T\langle\varphi_{T}\rangle=\omega^2\langle\varphi_{T}\rangle$ and $T\langle\phi\rangle=\omega^2\langle\phi\rangle$. However, the lepton
flavor mixing is associated with the hermitian combination $m_lm^{\dagger}_l$, which is obviously invariant under the action of $T$, i.e., $T^{\dagger}m_{l}m^{\dagger}_lT=m_{l}m^{\dagger}_l$. Consequently there is still a remnant $Z^{T}_3$ symmetry in the charged lepton sector if we concentrate on lepton flavor mixing. Furthermore, we can check that only three of the 24 generalised CP symmetries are preserved by $m_{l}m^{\dagger}_l$ and $H^{l}_{CP}=\left\{\rho_{\mathbf{r}}(1), \rho_{\mathbf{r}}(T), \rho_{\mathbf{r}}(T^2)\right\}$.


Neutrino masses are generated by type I see-saw mechanism. The LO superpotential is given by
\begin{equation}
\label{eq:linear-four}
w_{\nu}=\frac{y_1}{\Lambda}\left(l\nu^c\right)_{\mathbf{1}}\xi h_u
+\frac{y_2}{\Lambda}\left(\left(l\nu^c\right)_{\mathbf{2}}\eta\right)_{\mathbf{1}}h_u
+\frac{y_3}{\Lambda}\left(\left(l\nu^c\right)_{\mathbf{3}}\varphi_S\right)_{\mathbf{1}}h_u
+M\left(\nu^c\nu^c\right)_{\mathbf{1}}\,,
\end{equation}
where the first three terms contribute to the neutrino Dirac mass whereas the last one is the Majorana mass terms for the right-handed neutrinos.
All the couplings are again real because of the imposed generalised CP symmetry. Given the vacuum configuration of Eq.~\eqref{eq:neutrino-vacuum-two}, we can read out the Dirac and Majorana mass matrices as follows
\begin{eqnarray}
 \hskip-0.3in m_D=y_1v_u\frac{v_\xi}{\Lambda}\left[\left(\begin{array}{ccc}
    1  &  0  &  0  \\
    0  &  0  &  1  \\
    0  &  1  &  0
  \end{array}\right)+x\left(\begin{array}{ccc}
    0  &  1  &  1  \\
    1  &  1  &  0  \\
    1  &  0  &  1
  \end{array}\right)+y\left(\begin{array}{ccc}
   0  &  -1  &  1  \\
   1  &  0  &  2  \\
   -1  &  -2  &  0
 \end{array}\right)\right],\quad
  m_M=M\left(\begin{array}{ccc}
    1  &  0  &  0  \\
    0  &  0  &  1  \\
    0  &  1  &  0
  \end{array}\right)\,,
\end{eqnarray}
where $v_{u}=\langle h_u\rangle$ and the parameters $x$, $y$ are
\begin{equation}
x=\frac{y_2v_{\eta}}{y_1v_{\xi}},\qquad y=\frac{y_3v_{S}}{y_1v_{\xi}}.
\end{equation}
After extracting the common phase of the VEVs $v_S$, $v_{\eta}$ and $v_{\xi}$, the parameter $x$ is real, while $y$ is real or purely imaginary. The light neutrino mass matrix is given by the see-saw formula
\begin{eqnarray}
\nonumber m_{\nu}&=&-m_Dm^{-1}_Mm^T_D\\
\label{eq:neutrino_matrix_model1}  &=&\alpha \left(\begin{array}{ccc}
   2  &  -1  &  -1  \\
   -1  &  2  &  -1  \\
   -1  &  -1  &  2
 \end{array}\right)+\beta \left(\begin{array}{ccc}
    1  &  0  &  0  \\
    0  &  0  &  1  \\
    0  &  1  &  0
  \end{array}\right)+\gamma \left(\begin{array}{ccc}
    0  &  1  &  1  \\
    1  &  1  &  0  \\
    1  &  0  &  1
  \end{array}\right)+\delta \left(\begin{array}{ccc}
    0  &  1  &  -1  \\
    1  &  2  &  0  \\
    -1  &  0  &  -2
  \end{array}\right)\,.
\end{eqnarray}
It is the most general neutrino mass matrix consistent with the residual $Z^{SU}_2$ flavor symmetry, as is shown in Eq.~\eqref{eq:general-mass-matrix}. The four parameters $\alpha $, $\beta $, $\gamma $ and $\delta $ are given by
\begin{equation}
\alpha =-y^2m_0,~~\beta =(4y^2-2x^2-1)m_0,~~ \gamma =(y^2-x^2-2x)m_0,~~\delta =-3xym_0\,,
\end{equation}
where $m_0=\frac{y_1^2v_u^2v_\xi^2}{M\Lambda^2}$ is the overall scale of the light neutrino masses. We see that $\alpha$, $\beta$ and $\gamma$ are real parameters, $\delta$ is real or imaginary for $v_S$ being real or imaginary, respectively.
Furthermore, the effective mass parameter $|m_{\beta\beta}|$ for the neutrinoless double-beta decay is given by
\begin{equation}
\left|m_{\beta\beta}\right|=m_0\left|2\alpha+\beta\right|
\end{equation}

As shown in Eq.~\eqref{eq:VEV_relation}, if the combination $g_3g_4\left(g_3g_5+g_2g_6\right)>0$, which leads to real $v_S$ and $\delta$ parameters, the vacuum alignments of the flavons $\varphi_S$, $\eta$ and $\xi$ in Eq.~\eqref{eq:neutrino-vacuum-two} are invariant under the action of both $\rho_{\mathbf{r}}(1)$ and $\rho_{\mathbf{r}}(SU)$ elements of $H_{CP}$. Therefore the generalised CP symmetry is broken to $H^{\nu}_{CP}=\left\{\rho_{\mathbf{r}}(1),\rho_{\mathbf{r}}(SU)\right\}$ in the neutrino sector. This case is identical to case (I) of the general analysis inspired by symmetry arguments. The corresponding light neutrino mass matrix of Eq.~\eqref{eq:neutrino_matrix_model1} is real, the lepton mixing is exactly the $\text{TM}_1$ pattern with conserved CP, and the predictions for light neutrino masses and mixing angles are given in Eq.~\eqref{eq:mass-one} and Eq.~\eqref{eq:mixing-angles-one}. Notice that the light neutrino mass matrix of Eq.~\eqref{eq:neutrino_matrix_model1} depends on three real parameters $x$, $y$ and $m_0$, their values can be fixed by the measured values of the mass squared differences $\Delta m^2_{sol}$ and $\Delta m^2_{atm}$ and the reactor neutrino mixing angle $\theta_{13}$. As a result, both the absolute scale of the neutrino masses and the lepton mixing angles are fixed. For the best fitting values of $\Delta m^2_{\text{sol}}=7.45\times10^{-5}\text{eV}^2$, $\Delta m^2_{\text{atm}}=2.417(2.410)\times10^{-3}\text{eV}^2$ and $\sin^2\theta_{13}=0.0229$ from Ref.~\cite{GonzalezGarcia:2012sz},
we find there are 8 solutions to the values of $x$ and $y$ in the case that both $x$ and $y$ are real. The corresponding predictions for the light neutrino masses and the lepton mixing parameters are summarized in Table~\ref{tab:effective5}. 
It is obvious that the former 4 solutions correspond to a normal ordering (NO) neutrino mass spectrum, and the latter 4 correspond to inverted ordering (IO) spectrum. Moreover, we see that the predicted values for the atmospheric mixing angle $\theta_{23}$ ($32.496^{\circ}$ and $57.504^{\circ}$) are slightly beyond the $3\sigma$ range of the current global data fitting~\cite{Tortola:2012te,Fogli:2012ua,GonzalezGarcia:2012sz}. We note that the NLO corrections and the renormalization group evolution effects could bring the model to agree with the experimental data. However, in these scenarios a value of $\theta_{23}$ very close to the maximal mixing value of $45^{\circ}$ would be unnatural. The next generation neutrino oscillation
experiments, in particular those exploiting a high intensity neutrino beam, will reduce the experimental error on $\theta_{23}$ to few degrees. If no significant deviations from maximal atmospheric mixing will be detected, these 8 solutions will be ruled out.


\begin{table}[t!]
 \centering
 \renewcommand{\tabcolsep}{1.7mm}
 {\footnotesize
 \begin{tabular}{|c|c|c|c|c|c|c|c|c|c|c|}
 \hline  \hline
 $\left(x,y\right)$ & $\delta_{\rm{CP}}$ & $\theta_{23}/^\circ$  & $\theta_{12}/^\circ$  & $\alpha_{21}$ & $\alpha_{31}$ & $m_1$ & $m_2$   & $m_3$ & $|m_{\beta\beta}|$ & \text{mass order}\\ \hline
  $(-1.898,-0.316)$ & $\pi $ & 32.496  & \multirow{10}{*}{34.309}  & \multirow{2}{*}{0} & \multirow{2}{*}{$\pi$} & \multirow{2}{*}{128.020} & \multirow{2}{*}{128.311} & \multirow{2}{*}{137.136} & \multirow{2}{*}{122.038} & \multirow{2}{*}{NO} \\ \cline{1-3}
   $(-1.898,0.316)$ & 0 & 57.504 & & & & & & & & \\ \cline{1-3} \cline{5-11}
   $(0.139,-0.612)$ & $\pi$  & 32.496 &  & \multirow{2}{*}{$\pi $} & \multirow{2}{*}{0} & \multirow{2}{*}{24.233} & \multirow{2}{*}{25.724} & \multirow{2}{*}{54.747} & \multirow{2}{*}{9.423} & \multirow{2}{*}{NO} \\ \cline{1-3}
  $(0.139,0.612)$ & $0$ & 57.504 & & & & & & & & \\ \cline{1-3} \cline{5-11}
   $(0.101,0.340)$ & $\pi $ & 32.496 &   & \multirow{2}{*}{0} & \multirow{2}{*}{$\pi$} & \multirow{2}{*}{49.669} & \multirow{2}{*}{50.414} & \multirow{2}{*}{11.159} & \multirow{2}{*}{48.507} & \multirow{2}{*}{IO} \\ \cline{1-3}
   $(0.101,-0.340)$ & $0$ & 57.504 & & & & & & & & \\ \cline{1-3} \cline{5-11}
    $(-0.120,0.535)$ & $\pi$ & 32.496  &   & \multirow{2}{*}{$\pi $} & \multirow{2}{*}{0} & \multirow{2}{*}{54.866} & \multirow{2}{*}{55.541} & \multirow{2}{*}{25.977} & \multirow{2}{*}{19.931} & \multirow{2}{*}{IO} \\ \cline{1-3}
   $(-0.120,-0.535)$ & 0 & 57.504 & & & & & & & & \\ \cline{1-3} \cline{5-11}
    $(-0.050,0.233i)$ & $\pi/2$ & \multirow{2}{*}{45} &  & \multirow{2}{*}{0} & \multirow{2}{*}{0}  & \multirow{2}{*}{57.284} & \multirow{2}{*}{57.930} & \multirow{2}{*}{75.488} & \multirow{2}{*}{57.901} & \multirow{2}{*}{NO} \\ \cline{1-2}
   $(-0.050,-0.233i)$ & $-\pi/2$ & & & & & & & & & \\ \hline \hline
 \end{tabular}}
 \caption{\label{tab:effective5}The predictions for the leptonic CP phases, light neutrino masses $m_i(i=1,2,3)$  and the effective mass $|m_{\beta\beta}|$ of the neutrinoless doublet-beta decay, where the unit of mass is meV.}
\end{table}

Another possibility of $g_3g_4\left(g_3g_5+g_2g_6\right)<0$ gives rise to an imaginary $v_S$ such that the parameter $\delta$ in the neutrino mass matrix of Eq.~\eqref{eq:neutrino_matrix_model1} is purely imaginary as well. The remnant CP symmetry in the neutrino sector is $H^{\nu}_{CP}=\left\{\rho_{\mathbf{r}}(S), \rho_{\mathbf{r}}(U)\right\}$. This corresponds to the case (II) discussed in the general analysis of section~\ref{subsec:general_analysis}. The predictions for the mixing parameters and the light neutrino masses are given in Eq.~\eqref{eq:mixing-angles-two} and Eq.~\eqref{eq:mass-two}. The lepton mixing is of the $\text{TM}_1$ form, and maximal Dirac CP violation $\left|\delta_{CP}\right|=\pi/2$ and maximal atmospheric mixing $\theta_{23}=45^{\circ}$ are produced in this case. Analogously, the light neutrino sector is also controlled by three real parameters, and hence the model is quite predictive, as shown in the last two lines of Table~\ref{tab:effective5}. The neutrino mass spectrum can only be normal ordering in this case.

Generally the LO results are modified by the subleading terms invariant under the imposed symmetry. Because of the auxiliary $Z_7$ symmetry in the present model, all the subleading corrections can be obtained by inserting the combination $\Phi_{l}\Phi_{\nu}$ into the LO terms of $w_d$, $w_{l}$ and $w_{\nu}$ in Eqs.(\ref{eq:wd_model1}, \ref{eq:linear-one}, \ref{eq:linear-four})~\footnote{All possible $S_4$ contractions should be considered here, and only the correction to the electron mass terms is an exception with the form $\left(l\Phi^4_{\nu}\right)e^{c}h_d/\Lambda^4$.}, where $\Phi_{l}=\left\{\varphi_T,\phi\right\}$ and $\Phi_{\nu}=\left\{\varphi_S,\eta,\xi\right\}$ denote the flavons in the charged lepton and neutrino sectors respectively. As a result, the corresponding corrections to the lepton masses and mixing angles are suppressed by $\langle\Phi_l\rangle\langle\Phi_{\nu}\rangle/\Lambda^2\sim\lambda^4$ with respect to the LO contributions and therefore can be negligible.

\section{\label{sec:model_majorana}Model 2}

\cleqn

\begin{table} [h]
\centering
\renewcommand{\tabcolsep}{1.8mm}
\footnotesize{
\begin{tabular}{|c||c|c|c|c|c|c||c|c|c|c|c|c||c|c|c|c|c|c|c|}
\hline \hline
Field &   $l$    &  $\nu^c$     &  $e^{c}$     &   $\mu^c$    &    $\tau^c$  &  $h_{u,d}$   &  $\varphi_T$ & $\phi $  &   $\varphi_S$  &    $\eta$  &  $\chi $ &  $\xi $   &  $\varphi^{0}_T$  &  $\zeta^0$  &  $\varphi^{0}_S$   &  $\xi^{0}$ & $\eta ^0$ &  $\rho ^0$ &  $\sigma ^0$ \\ \hline

$S_4$  &   $\mathbf{3}^\prime$  &  $\mathbf{3}^\prime$  &  $\mathbf{1}$  & $\mathbf{1}$   &  $\mathbf{1}$   & $\mathbf{1}$   &   $\mathbf{3}^\prime$ & $\mathbf{2}$ &  $\mathbf{3}^\prime$  &   $\mathbf{2}$ &  $\mathbf{3^\prime}$ & $\mathbf{1}$   &  $\mathbf{3}'$  &  $\mathbf{1}$ &  $\mathbf{3}$   &   $\mathbf{1}$ & $\mathbf{2}$ &   $\mathbf{1}$ &  $1$ \\ \hline

$Z_4$  &    $1$  &   $1$    &    $i$    &   $-1$   &    $-i$  &  $1$   &  $i$ & $i$  &  $1$   &   $1$   & $1$  & $1$ &  $-1$  &  $-1$  &  $1$   &  $1$ &  $1$ &  $1$ & $1$ \\ \hline

$Z_5$  & $\omega^3_5$ &  $\omega^2_5$ &  $\omega^2_5$ & $\omega^2_5$ & $\omega^2_5$ & $1$  & $1$ & $1$ & $\omega_5$ & $\omega_5$ & $\omega^3_5$  & $\omega^3_5$ & $1$ & $1$ & $\omega^3_5$ & $\omega^3_5$  & $\omega^4_5$ &  $\omega^4_5$ &  $\omega_5$ \\ \hline

$U(1)_R$  &  $1$  &   $1$  &   $1$  &   $1$  &   $1$  &  $0$ &    $0$  &  $0$ & $0$  & $0$  &   $0$ & $0$  &  $2$ &   $2$ &   $2$ &   $2$ & $2$ &$2$ & $2$ \\ \hline \hline
\end{tabular}}
\caption{\label{tab:effective-one}The transformation properties of the fields under the family symmetry $S_4\times Z_4\times Z_5$ and $U(1)_R$, where $\omega_5=e^{\frac{2\pi i}{5}}$.}
\end{table}

In this section, we shall try to improve the previous model by generating the reactor mixing angle at the next-to-leading order (NLO) such that the correct order of magnitude of $\theta_{13}$ is produced. In this model, the LO lepton mixing is the well-known tri-bimaximal mixing pattern which is broken to $\text{TM}_1$ mixing by NLO contributions. Analogous to the previous model, the present model is based on the symmetry $S_4\rtimes H_{CP}$ with the extra symmetry $Z_4\times Z_5\times U(1)_R$ in order to eliminate unwanted operators. The matter fields, flavon fields, driving fields and their transformation rules under the family symmetry are summarized in Table~\ref{tab:effective-one}. As previous model of section~\ref{sec:model_dirac}, the remnant symmetry of the hermitian combination $m_lm^{\dagger}_l$ is $Z^{T}_3\rtimes H^{l}_{CP}$ with $H^{l}_{CP}=\left\{\rho_{\mathbf{r}}(1), \rho_{\mathbf{r}}(T), \rho_{\mathbf{r}}(T^2)\right\}$, and the original symmetry $S_4\rtimes H_{CP}$ is broken down to $G^{nu}_{CP}=Z^{SU}_2\times H^{\nu}_{CP}$. As a consequence, the model-independent analysis results of section~\ref{subsec:general_analysis} are realized within one model, and the Dirac CP phase $\delta_{CP}$ is predicted to be trivial or maximal. In the following, we firstly discuss the vacuum alignment of the model, then specify the structure of the model at LO and NLO.

\subsection{\label{sec:vacuum_alignment_model2}Vacuum alignment}
The most general driving superpotential $w^{l}_d$ associated with the charged lepton sector, which is invariant under the family symmetry $S_4\times Z_4\times Z_5$, can be written as
\begin{equation}
\label{eq:driving-potential-six}
w^l_d=f_1\left(\varphi_T^0\left(\varphi_T\varphi_T\right)_{\mathbf{3}^{\prime}}\right)_{\mathbf{1}}
+f_2\left(\varphi_T^0\left(\phi\varphi_T\right)_{\mathbf{3}^{\prime}}\right)_{\mathbf{1}}+f_3\zeta^0\left(\varphi_T\varphi_T\right)_{\mathbf{1}}+f_4\zeta^0\left(\phi\phi\right)_{\mathbf{1}}\,.
\end{equation}
It is exactly the same as Eq.~\eqref{eq:driving-potential-one}. Hence the vacuum of the flavon fields $\varphi_T$ and $\phi$ is of the same form as shown in Eq.~\eqref{eq:charged-vacuum-one}, i.e.
\begin{eqnarray}
\label{eq:charged-vacuum-two}
\langle\varphi_T\rangle=\left(\begin{array}{c}
0\\
1\\
0
\end{array}\right)v_T,\qquad
\langle\phi \rangle=\left(\begin{array}{c}
0\\
1
\end{array}\right)v_{\phi},\quad \text{with} ~~~v_T=-\frac{f_2}{2f_1}v_{\phi}\,.
\end{eqnarray}
We see that $v_{\phi}$ and $v_T$ carry the same phase up to $\pi$. Since the phase of $v_{\phi}$ can be absorbed by leptons, we can take $v_\phi$ and $v_T$ to be real without loss of generality.
From the following predictions for charged lepton masses in Eq.~\eqref{eq:charged_lepton_mass_model2}, we note that the mass hierarchies between the charged leptons can be produced for
\begin{eqnarray}
\label{eq:VEVs_order_CH_model2}\frac{v_{\phi}}{\Lambda}\sim\frac{v_T}{\Lambda}\sim\mathcal{O}(\lambda^2)\,.
\end{eqnarray}
The driving superpotential $w^{\nu}_d$ involving the flavons of the neutrino sector reads
\begin{eqnarray}
\label{eq:driving-potential-two}
\nonumber w^{\nu}_d&=&g_1\left(\varphi_S^0\left(\varphi_S\varphi_S\right)_{\mathbf{3}}\right)_{\mathbf{1}}
+g_2\left(\varphi_S^0\left(\eta\varphi_S\right)_{\mathbf{3}}\right)_{\mathbf{1}}
+g_3\xi^0\left(\varphi_S\varphi_S\right)_{\mathbf{1}}+g_4\xi^0\left(\eta\eta\right)_{\mathbf{1}}+M_{\eta}\left(\eta^0\eta\right)_{\mathbf{1}}\\
&&\hskip-0.1in+g_5\left(\eta^0\left(\chi\chi\right)_{\mathbf{2}}\right)_{\mathbf{1}}+g_6\rho^0\left(\chi\chi\right)_{\mathbf{1}}
+g_7\rho^0\xi^2+g_8\sigma^0\left(\chi\varphi_S\right)_{\mathbf{1}}\,,
\end{eqnarray}
where all the coupling $g_i$ and mass parameter $M_{\eta}$ are real because of the imposed generalised CP symmetry. Since the contraction $\left(\varphi_S\varphi_S\right)_{\mathbf{3}}$ vanishes due to the antisymmetry of the associated $S_4$ Clebsch-Gordan coefficients, the first term proportional to $g_1$ gives null contribution. In the SUSY limit, the vacuum configuration is determined by the vanishing of the derivative of the driving superpotential $w^{\nu}_d$ with respect to each component of the driving fields. The minimization equations for the vacuum take the following form:
\begin{eqnarray}
\nonumber&&\frac{\partial w^{\nu}_d}{\partial\varphi^0_{S_1}}=
g_2\left(\eta_{1}\varphi_{{S_2}}-\eta_{2}\varphi_{{S_3}}\right)=0\,, \\
\nonumber&&\frac{\partial w^{\nu}_d}{\partial\varphi^0_{S_2}}=
g_2\left(\eta_{1}\varphi_{S_1}-\eta_{2}\varphi_{S_2}\right)=0\,, \\
\nonumber&&\frac{\partial w^{\nu}_d}{\partial\varphi_{S_3}^0}=
g_2\left(\eta_{1}\varphi_{S_3}-\eta_{2}\varphi_{S_1}\right)=0\,,  \\
\nonumber&&\frac{\partial w^{\nu}_d}{\partial\xi^0}=g_3\left(\varphi_{S_1}^2+2\varphi_{S_2}\varphi_{S_3}\right)+2g_4\eta_{1}\eta_{2}=0\,,\\
\nonumber&&\frac{\partial w_d}{\partial\eta^0_1}=M_{\eta}\eta_{2}+g_5\left(\chi^2_{3}+2\chi_{1}\chi_{2}\right)=0\,,\\
\nonumber&&\frac{\partial w_d}{\partial\eta^0_2}=M_{\eta}\eta_{1}+g_5\left(\chi^2_{2}+2\chi_{1}\chi_{3}\right)=0\,,
\end{eqnarray}
\begin{eqnarray}
\nonumber&&\frac{\partial w_d}{\partial\rho^0}=g_6\left(\chi_{1}^2+2\chi_{2}\chi_{3}\right)+g_7\xi^2=0\,,\\
&&\frac{\partial w_d}{\partial\sigma^0}=g_8\left(\chi_1\varphi_{S_1}+\chi_2\varphi_{S_3}+\chi_3\varphi_{S_2}\right)=0\,.
\end{eqnarray}
The solution to these equation are
\begin{equation}
\label{eq:neutrino-vacuum-one}
\langle\varphi_S\rangle =\left(\begin{array}{c}
   1 \\
   1  \\
   1
  \end{array}\right)v_S, \quad
\langle\eta\rangle=\left(\begin{array}{c}
    1 \\
    1
\end{array}\right)v_{\eta} ,\quad
\langle\chi\rangle=\left(\begin{array}{c}
 0 \\
 1  \\
 -1
\end{array}\right)v_{\chi},\quad \langle\xi\rangle=v_{\xi} \,.
\end{equation}
The VEVs $v_S$, $v_{\eta}$, $v_{\chi}$ and $v_{\xi}$ are related by
\begin{equation}
\label{eq:VEVs_relation_model2}v^2_S=-\frac{g_4g^2_5g^2_7}{6g_3g^2_6M^2_{\eta}}v^4_{\xi},\qquad v_{\eta}=-\frac{g_5g_7}{2g_6M_{\eta}}v^2_{\xi},\qquad v^2_{\chi}=\frac{g_7}{2g_6}v^2_{\xi}\,,
\end{equation}
where $v_{\xi}$ parameterizes a flat direction in the driving superpotential $w^{\nu}_d$, and it is in general complex. It is straightforward to check that the VEVs of the flavon fields $\varphi_S$, $\eta$ and $\xi$ preserve the remnant $K_4$ subgroup generated by $Z_2^{S}$ and $Z_2^{SU}$, while the VEV of $\chi$ is invariant only under the action of $Z_2^{SU}$. In our model presented below, $\varphi_S$ and $\eta$ couple with the right-handed neutrino at LO, as shown in Eq.~\eqref{eq:neutrino_yukawa_model2}. The resulting lepton mixing is of the tri-bimaximal form. The flavons $\chi$ and $\xi$ enter into the neutrino sector at NLO, and the LO residual $K_4$ symmetry is further broken down to $Z_2^{SU}$. As a result, the NLO contributions modify the LO tri-bimaximal mixing into $\text{TM}_1$ pattern. In order to achieve the measured size of $\theta_{13}\simeq\lambda/\sqrt{2}$~\cite{King:2012vj,Minakata:2004xt}, we could choose
\begin{eqnarray}
\label{eq:hierarchy-one}
\frac{v_S}{\Lambda}\sim\frac{v_{\eta}}{\Lambda}\sim\frac{v_{\chi}}{\Lambda}\sim\frac{v_{\xi}}{\Lambda}\sim \mathcal{O}(\lambda)\,.
\end{eqnarray}
Consequently the NLO corrections are suppressed by a factor $\lambda$ with respect to the LO contributions, and therefore the reactor angle is of the correct order $\lambda$. Note that the VEVs of the flavon fields in the neutrino and the charged lepton sectors are chosen to be of different order of magnitude: $\lambda\Lambda$ v.s. $\lambda^2\Lambda$, please see Eq.~\eqref{eq:VEVs_order_CH_model2} and Eq.~\eqref{eq:hierarchy-one}. This mild hierarchy can be accommodated because these two sets of VEVs depend on  different model parameters.

\subsection{\label{sec:LO_model2} Leading order results}

The superpotential for the charged lepton masses, which is allowed by the symmetry, is given by
\begin{eqnarray}
\label{eq:linear-three}
\nonumber w_l&&=\frac{y_{\tau}}{\Lambda}\left(l\varphi_T\right)_{\mathbf{1}}\tau^{c}h_d+\frac{y_{\mu_1}}{\Lambda^2} \left(l\left(\varphi_T\varphi_T\right)_{\mathbf{3}^{\prime}}\right)_{\mathbf{1}}\mu^{c}h_d
+\frac{y_{\mu_2}}{\Lambda^2}\left(l\left(\phi\varphi_T\right)_{\mathbf{3}^{\prime}}\right)_{\mathbf{1}}\mu^{c}h_d\\
\nonumber&&+\frac{y_{e_1}}{\Lambda^3}\left(l\varphi_T\right)_{\mathbf{1}}\left(\varphi_T\varphi_T\right)_{\mathbf{1}}e^{c}h_d
+\frac{y_{e_2}}{\Lambda^3}\left(\left(l\varphi_T\right)_{\mathbf{2}}\left(\varphi_T\varphi_T\right)_{\mathbf{2}}\right)_{\mathbf{1}}e^ch_d
+\frac{y_{e3}}{\Lambda^3}\left(\left(l\varphi_T\right)_{\mathbf{3}^{\prime}}\left(\varphi_T\varphi_T\right)_{\mathbf{3}^{\prime}}\right)_{\mathbf{1}}e^ch_d\\
\nonumber&&
+\frac{y_{e_4}}{\Lambda^3}\left(\left(l\varphi_T\right)_{\mathbf{3}}\left(\varphi_T\varphi_T\right)_{\mathbf{3}}\right)_{\mathbf{1}}e^ch_d
+\frac{y_{e_5}}{\Lambda^3}\left(\left(l\phi\right)_{\mathbf{3}^{\prime}}\left(\varphi_T\varphi_T\right)_{\mathbf{3}^{\prime}}\right)_{\mathbf{1}}e^ch_d  +\frac{y_{e_6}}{\Lambda^3}\left(\left(l\phi\right)_{\mathbf{3}}\left(\varphi_T\varphi_T\right)_{\mathbf{3}}\right)_{\mathbf{1}}e^ch_d\\
&&+\frac{y_{e_7}}{\Lambda^3}\left(\left(l\varphi_T\right)_{\mathbf{2}}\left(\phi\phi\right)_{\mathbf{2}}\right)_{\mathbf{1}}e^ch_d +\frac{y_{e_8}}{\Lambda^3}\left(l\varphi_T\right)_{\mathbf{1}}\left(\phi\phi\right)_{\mathbf{1}}e^ch_d+\ldots\,,
\end{eqnarray}
which is identical to the corresponding superpotential of Model 1 shown in  Eq.~\eqref{eq:linear-one}. After electroweak and flavor symmetry breaking in the way of Eq.\eqref{eq:charged-vacuum-two}, we obtain a diagonal charged lepton mass matrix:
\begin{eqnarray}
\label{eq:charged_lepton_mass_model2}
m_l=\left( \begin{array}{ccc}
y_e\frac{v_T^2}{\Lambda^2} & 0  & 0 \\
0  &  y_{\mu}\frac{v_T}{\Lambda} &  0 \\
0  &  0  &  y_\tau
\end{array}\right)\frac{v_T}{\Lambda}v_d\,,
\end{eqnarray}
where $y_{e}$ and $y_{\mu}$ are the results of the different contributions of the $y_{e_i}$ and $y_{\mu_{i}}$ respectively with
\begin{equation}
y_{\mu}=2y_{{\mu}_1}+y_{{\mu}_2}\frac{v_{\phi}}{v_T},\qquad  y_e=y_{e2}-2y_{e_3}+2y_{e_5}\frac{v_{\phi}}{v_T}+y_{e_7}\frac{v^2_{\phi}}{v^2_{T}}\,.
\end{equation}
Now we turn to the neutrino sector, The LO superpotential relevant to the neutrino masses is of the form
\begin{equation}
\label{eq:neutrino_yukawa_model2}
w_{\nu}=y\left(l\nu^c\right)_{\mathbf{1}}h_u+y_1\left(\left(\nu^c\nu^c\right)_{\mathbf{3}^{\prime}}\varphi_S\right)_{\mathbf{1}}+y_2 \left(\left(\nu^c\nu^c\right)_{\mathbf{2}}\eta\right)_{\mathbf{1}}\,,
\end{equation}
where all the three couplings $y$, $y_1$ and $y_2$ are real because of the generalised CP symmetry. We can easily read out the Dirac neutrino mass matrix as
\begin{equation}
m_D=yv_u \left(\begin{array}{ccc}
   1  &  0  &  0  \\
   0  &  0  &  1  \\
   0  &  1  &  0
\end{array}\right)\,.
\end{equation}
Given the vacuum of the flavons $\varphi_S$ and $\eta$ shown in Eq.~\eqref{eq:neutrino-vacuum-one}, the mass matrix for the right-handed neutrino takes the form
\begin{eqnarray}
m_M=a\left( \begin{array}{ccc}
   2  &  -1  &  -1  \\
   -1  &  2  &  -1  \\
   -1  &  -1  &  2
\end{array}\right)
+b\left(\begin{array}{ccc}
    0  & 1 & 1 \\
    1  & 1 & 0 \\
    1  & 0 & 1
\end{array} \right)\,,
\end{eqnarray}
where $a=y_1v_{S}$ and $b=y_2v_{\eta}$. The light neutrino mass matrix is given by the see-saw formula, yielding
\begin{eqnarray}
\label{eq:neutrino-mass-matrix1}
m_\nu=-m_Dm_M^{-1}m_D^T=U_{TB}\text{diag}(m_1,m_2,m_3)U_{TB}^T\,.
\end{eqnarray}
That is to say the LO lepton flavor mixing is the tri-bimaximal pattern. The reason is that the VEVs of $\varphi_S$ and $\eta$ break the $S_4$ family symmetry into a residual $K_4\cong Z^{S}_2\times Z^{SU}_2$ subgroup, i.e. the vacuum of $\varphi_S$ and $\eta$ in Eq.~\eqref{eq:neutrino-vacuum-one} is invariant under both $Z^{S}_2$ and $Z^{SU}_2$. Furthermore, the light neutrino masses $m_{1,2,3}$ in Eq.~\eqref{eq:neutrino-mass-matrix1} are given by
\begin{eqnarray}
\label{eq:neutrino-mass-one}
m_1=\frac{y^2v^2_u}{-3y_1v_S+y_2v_{\eta}}, \qquad m_2=-\frac{y^2v^2_u}{2y_2v_{\eta}}, \qquad m_3=-\frac{y^2v^2_u}{3y_1v_S+y_2v_{\eta}}\,.
\end{eqnarray}
It is interesting to note that the following sum rule is satisfied
\begin{equation}
\frac{1}{m_3}-\frac{1}{m_1}=\frac{1}{m_2}\,.
\end{equation}
Since the VEVs $v_{S}$ and $v_{\eta}$ are related through Eq.~\eqref{eq:VEVs_relation_model2}, the phase different between $v_S$ and $v_{\eta}$ is fixed to discrete values $0$, $\pi$ or $\pm\pi/2$ for the product $g_3g_4<0$ or $g_3g_4>0$, respectively. Moreover, the phase of $v_\xi $ can be absorbed by redefining the right-handed neutrino fields, therefore we can set $v_{\xi}$ to be real, and then another VEV $v_S$ would be real or purely imaginary. For the case of $v_S$ being imaginary, Eq.~\eqref{eq:neutrino-mass-one} implies that the light neutrino masses are degenerate, i.e. $|m_1|=|m_3|$. Therefore this case is not phenomenologically viable, and we shall choose $v_S$ to be real (or $v_{S}$ and $v_{\eta}$ have the same phase up to relative sign) in the following. Then the neutrino mass-squared differences are predicted to be
\begin{eqnarray}
\nonumber&&\Delta m_{sol}^2\equiv |m_2|^2-|m_1|^2=\frac{3(3x+1)(x-1)}{4(3x-1)^2}\left(\frac{y^2v^2_u}{y_2v_{\eta}}\right)^2 \,, \\
\nonumber&&\Delta m_{atm}^2\equiv |m_3|^2-|m_1|^2=\frac{-12x}{(9x^2-1)^2}\left(\frac{y^2v^2_u}{y_2v_{\eta}}\right)^2, \quad \text{for} \quad \text{NO}\,, \\
\label{eq:mass_squared_difference_model2}&& \Delta m_{atm}^2\equiv |m_2|^2-|m_3|^2=\frac{3(3x-1)(x+1)}{4(3x+1)^2}\left(\frac{y^2v^2_u}{y_2v_{\eta}}\right)^2, \quad \text{for} \quad \text{IO}\,,
\end{eqnarray}
where $x=\frac{y_1v_S}{y_2v_{\eta}}$ is a real parameter. Furthermore, the effective mass parameter $|m_{\beta\beta}|$ for the neutrinoless doublet beta is given by
\begin{equation}
|m_{\beta\beta }|=\left|\frac{x+1}{2(3x-1)}\right|\left|\frac{y^2v^2_u}{y_2v_{\eta}}\right|\,.
\end{equation}
Since the solar neutrino mass squared difference $\Delta m_{sol}^2$ is positive, we have $x>1$ or $x<-\frac{1}{3}$ from Eq.~\eqref{eq:mass_squared_difference_model2}. By further inspecting the atmospheric neutrino mass squared difference $\Delta m^2_{atm}$, we find that neutrino spectrum is normal ordering (NO) for $x<-\frac{1}{3}$ and inverted order (IO) for $x>1$. Taking the best fit values $\Delta m^2_{sol}=7.45\times 10^{-5}\text{eV}^2$ and $\Delta m^2_{atm}=2.417(2.410)\times 10^{-3}\text{eV}^2$ for NO (IO) spectrum from Ref.~\cite{GonzalezGarcia:2012sz}, we get two solutions for the ratio $x$ (one for normal ordering and another for inverted ordering):
\begin{equation}
 x=-0.5173,\quad 1.0079\,.
\end{equation}
The corresponding predictions for the Majorana phases, the light neutrino masses and $|m_{\beta\beta}|$ are presented in Table~\ref{tab:effective6}.

\begin{table} [t!]
\centering
\begin{tabular}{|c|c|c|c|c|c|c|c|}
\hline\hline
$x$ &     $\alpha _{21}$    &  $\alpha _{31}$     &  $|m_1|$(meV)    &    $|m_2|$(meV)    &     $|m_3|$(meV)   &  $|m_{\beta\beta}|$(meV)  &   mass order  \\ \hline

$-0.5173$  &    $\pi$  &  0  &  10.891  &   13.897   &  50.355   & 2.628   &   NO \\ \hline

1.0079  &     0  &    0    &   55.913    &  56.576   &   28.121  & 56.134   &  IO  \\ \hline \hline
\end{tabular}
\caption{\label{tab:effective6}The predictions for the Majorana phases, the light neutrino masses $|m_i|$ ($i=1,2,3$) and the effective mass $|m_{\beta\beta}|$ of the neutrinoless double-beta decay at LO.}
\end{table}

\subsection{\label{sec:NLO_model2} Next-to-Leading-Order corrections}

Since the LO tri-bimaximal mixing pattern leads to a vanishing reactor angle $\theta_{13}$ which has been definitely excluded by the experimental measurements, NLO corrections are needed to achieve agreement with the present data. In this section, we shall address the NLO corrections indicated by higher dimensional operators compatible with all the symmetries of the model. As we shall show, the NLO contributions break the remnant family $K_4\cong Z^{S}_2\times Z^{SU}_2$ in the neutrino sector down to $Z^{SU}_2$. As a result, a non-zero $\theta_{13}$ is generated and it is naturally smaller than $\theta_{12}$ and $\theta_{23}$ which arise at LO.

In the following, we first discuss the NLO corrections to the charged lepton sector. For the driving superpotential $w^{l}_{d}$, the most relevant subleading operators can be written as
\begin{eqnarray}
\delta w^l_d&=&\left(\varphi^0_T\Psi^2_l\Psi^2_{\nu}\Psi^{\prime}_{\nu} \right)_{\mathbf{1}}/\Lambda^3+\left(\zeta^0\Psi^2_l\Psi^2_{\nu}\Psi^{\prime}_{\nu}\right)_{\mathbf{1}}/\Lambda^3\,,
\end{eqnarray}
where we have suppressed all dimensionless coupling constants, and all the possible $S_4$ contractions should be considered with $\Psi_l=\{\varphi_T, \phi\}$, $\Psi_{\nu}=\{\varphi_S, \eta\}$ and $\Psi^{\prime}_{\nu}=\{\chi,\xi\}$. These operators are suppressed by $\langle\Psi_{\nu}\rangle^2\langle\Psi^{\prime}_{\nu}\rangle/\Lambda^3\sim\lambda^3$ compared to LO terms in $w^{l}_d$ of Eq.~\eqref{eq:driving-potential-six}. Hence the subleading corrections to the VEVs of the $\varphi_T$ and $\phi$ appear at the relative order $\lambda^3$ such that their vacuum configurations at NLO can be parameterized as
\begin{equation}
\label{eq:vacuum_NLO_charged}
\langle\varphi_T\rangle =v_T\left(\begin{array}{c}
  \epsilon_1\lambda^3 \\
  1+\epsilon_2\lambda^3 \\
  \epsilon_3\lambda^3
\end{array}\right),\qquad
\langle\phi\rangle=v_{\phi}\left(\begin{array}{c}
   \epsilon_4\lambda^3 \\
   1
\end{array}\right)
\end{equation}
where the coefficients $\epsilon_i\;(i=1,2,3,4)$ have absolute value of order one and are generally complex due to the undetermined phase of $v_{\xi}$. Note that the shift of the second component of $\phi$ has been absorbed into the redefinition of the undetermined parameters $v_{\phi}$. The subleading corrections to the charged lepton superpotential $w_l$ take the form
\begin{eqnarray}
\label{eq:NLO-linear-one}
\delta w_l&=&\left(l\Psi_{l}\Psi^{2}_{\nu}\Psi^{\prime}_{\nu}\right)_{\mathbf{1}}h_d\tau^c/\Lambda^4
+\left(l\Psi^2_{l}\Psi^2_{\nu}\Psi^{\prime}_{\nu}\right)_{\mathbf{1}}h_d\mu^c/\Lambda^5
+\left(l\Psi^{3}_l\Psi^2_{\nu}\Psi^{\prime}_{\nu}\right)_{\mathbf{1}}h_de^c/\Lambda^6\,,
\end{eqnarray}
where the dimensionless coupling constants are omitted.
The charged lepton mass matrix is obtained by adding the contributions of this set of high dimensional operators evaluated with the insertion
of the LO VEVs of Eqs.~(\ref{eq:charged-vacuum-two},\ref{eq:neutrino-vacuum-one}), to those of the LO superpotential in Eq.~\eqref{eq:linear-three} evaluated
with the NLO vacuum configuration in Eq.~\eqref{eq:vacuum_NLO_charged}. We
find that each element of the charged lepton mass matrix receives corrections from both the subleading operators $\delta w_l$ in Eq.~\eqref{eq:NLO-linear-one} and the shifted vacuum alignment of Eq.~\eqref{eq:vacuum_NLO_charged}. As a consequence, its off-diagonal elements become non-zero and are all suppressed by $\lambda^3$ with respect to the diagonal ones. Therefore the charged lepton mass matrix including subleading corrections can be written as
\begin{eqnarray}
\label{eq:charged-mass-correction-two}
m^{NLO}_l=\left(\begin{array}{ccc}
m_e  &  \lambda^3 m_{\mu}  &  \lambda^3m_{\tau} \\
\lambda^3 m_e  &  m_{\mu}  &  \lambda^3m_{\tau} \\
\lambda^3 m_e  &  \lambda^3 m_{\mu} & m_{\tau}
\end{array}\right)\,.
\end{eqnarray}
Its contribution to the lepton mixing angles is of order $\lambda^3$ and can be safely neglected. Since the off-diagonal elements are quite small in particular the (2,1) and (3,1) entries, perturbatively diagonalizing the above NLO charged lepton mass matrix $m^{NLO}_l$ reveals that the NLO corrections to the charged lepton masses are of relative order $\lambda^6$, and hence they are negligible as well.

Next, we turn to discuss the NLO corrections in the neutrino sector. The NLO contributions to the driving superpotential $w^{\nu}_d$ is suppressed by one power of $1/\Lambda$ with respect to the LO terms in Eq.~\eqref{eq:driving-potential-two}, and it takes the form~\footnote{The subleading corrections to the terms proportional to $\eta^{0}$ and $\rho^{0}$ are of the form $\left(\eta^{0}\Psi^{3}_{\nu}\Psi^{\prime}_{\nu}\right)_{\mathbf{1}}/\Lambda^2$ and $\left(\rho^{0}\Psi^{3}_{\nu}\Psi^{\prime}_{\nu}\right)_{\mathbf{1}}/\Lambda^2$, which are suppressed by $1/\Lambda^2$ instead of $1/\Lambda$.}
\begin{eqnarray}
\label{eq:driving-potential-four}
\nonumber\delta w^{\nu}_d &=&\frac{h_1}{\Lambda}\left(\left(\varphi^0_S\varphi_S\right)_{\mathbf{2}}\left(\chi \chi\right)_{\mathbf{2}}\right)_{\mathbf{1}}
+\frac{h_2}{\Lambda}\left(\left(\varphi^0_S\varphi_S\right)_{\mathbf{3}}\left(\chi \chi\right)_{\mathbf{3}}\right)_{\mathbf{1}}
+\frac{h_3}{\Lambda}\left(\left(\varphi^0_S\varphi_S\right)_{\mathbf{3^{\prime}}}\left(\chi\chi\right)_{\mathbf{3^{\prime}}}\right)_{\mathbf{1}}\\
\nonumber&&+\frac{h_4}{\Lambda}\xi\left(\left(\varphi^0_S\varphi_S\right)_{\mathbf{3^{\prime}}}\chi\right)_{\mathbf{1}}
+\frac{h_5}{\Lambda}\left(\left(\varphi^0_S\eta\right)_{\mathbf{3}}\left(\chi\chi\right)_{\mathbf{3}}\right)_{\mathbf{1}}
+\frac{h_6}{\Lambda}\left(\left(\varphi^0_S\eta\right)_{\mathbf{3^{\prime}}}\left(\chi\chi\right)_{\mathbf{3^{\prime}}}\right)_{\mathbf{1}}\\
\nonumber&&+\frac{h_7}{\Lambda}\left(\left(\varphi^0_S\eta\right)_{\mathbf{3^{\prime}}}\chi\right)_{\mathbf{1}}\xi
+\frac{h_8}{\Lambda}\xi^0\left(\varphi_S\left(\chi\chi\right)_{\mathbf{3^{\prime}}}\right)_{\mathbf{1}}
+\frac{h_9}{\Lambda}\xi^0\xi\left(\varphi_S\chi\right)_{\mathbf{1}}\\
&&+\frac{h_{10}}{\Lambda}\xi^0\left(\eta\left(\chi\chi\right)_{\mathbf{2}}\right)_{\mathbf{1}}
+\frac{h_{11}}{\Lambda}\sigma^0\left(\chi\left(\chi\chi\right)_{\mathbf{3^{\prime}}}\right)_{\mathbf{1}}
+\frac{h_{12}}{\Lambda}\sigma^0\xi\left(\chi\chi\right)_{\mathbf{1}}+\frac{h_{13}}{\Lambda}\sigma^0\xi^3\,,
\end{eqnarray}
where all the couplings $h_i$ are again real because of the generalised CP symmetry. Repeating the minimization procedure of section~\ref{sec:vacuum_alignment_model2}, we find that the LO vacuum configuration is modified into
\begin{eqnarray}
\label{eq:vacuum_NLO_model2}
\langle\varphi_S\rangle=v^{\prime}_{S}\left(\begin{array}{c}
    1\\
    1\\
    1
\end{array}\right)+\delta v_S\left(\begin{array}{c}
    0\\
    1\\
    -1
\end{array}\right),\qquad
\langle\chi\rangle=v_{\chi}\left(\begin{array}{c}
    0\\
    1\\
    -1
\end{array}\right)+\delta v_{\chi}\left(\begin{array}{c}
    1\\
    1\\
    1
  \end{array}\right)\,,
\end{eqnarray}
with
\begin{eqnarray}
\nonumber&&v^{\prime}_S-v_{S}=-\left(\frac{h_8}{g_3}+\frac{h_{10}}{3g_3}\frac{v_{\eta}}{v_S}\right)\frac{v_{\chi}^2}{\Lambda}\,,\\
\nonumber&&\delta v_S=\left(\frac{h_4}{g_2}\frac{v_S}{v_{\eta}}-\frac{h_7}{g_2}\right)\frac{v_{\chi} v_{\xi}}{\Lambda}\,,\\
&&\delta v_{\chi}=-\frac{h_{13}v_{\xi}}{3g_8v_S}\frac{v^2_{\xi}}{\Lambda}
+\frac{2v_{\xi}}{3v_S}\left(\frac{h_{12}}{g_8}-\frac{h_7}{g_2}+\frac{h_4v_S}{g_2v_{\eta}}\right)\frac{v^2_{\chi}}{\Lambda}\,,
\end{eqnarray}
and the vacuum of $\eta$ doesn't acquires non-trivial shifts at this order. Obviously the shifts $v^{\prime}_S-v_S$, $\delta v_{S}$ and $\delta v_{\chi}$ are suppressed with respect to the LO VEVs $v_S$ and $v_{\chi}$ by a factor $\lambda$. Notice that the shifted vacuum of $\varphi_S$ and $\chi$ in Eq.~\eqref{eq:vacuum_NLO_model2} is the most general form of VEV invariant under the $Z^{SU}_2$ subgroup. 
The reason is that the NLO terms $\delta w^{\nu}_d$ of Eq.~\eqref{eq:driving-potential-four} only involve the neutrino flavons $\varphi_S$, $\eta$, $\chi$ and $\xi$ whose LO VEVs leave $Z^{SU}_2$ invariant.

From section~\ref{sec:LO_model2}, we know that the VEVs $v_S$, $v_{\eta}$ and $v^2_{\xi}$ have to share the same phase, i.e. the product $g_3g_4<0$ is needed otherwise the light neutrino mass spectrum would be partially degenerate. Furthermore, Eq.~\eqref{eq:neutrino-vacuum-one} implies that the phase different between $v_\chi$ and $v_\xi$ is $0$, $\pi$ or $\pm\frac{\pi}{2}$ for $g_6g_7>0$ or $g_6g_7<0$, respectively. As a result, $v^{\prime}_S$ and $v_S$ carry the sane phase. Since it is always possible to absorb the phase of $v_{\xi}$ by a redefinition of the matter fields, we
can take $v_{\xi}$ to be real without loss of generality. Then $v^\prime_S$, $v_{\eta}$ and $v^2_{\chi}$ would be real, while $v_{\chi}$ and $\delta v_{S}$ can be real or purely imaginary depending on $g_6g_7>0$ or $g_6g_7<0$.

Now we come to the NLO corrections to the LO neutrino superpotential $w_{\nu}$ in Eq.~\eqref{eq:neutrino_yukawa_model2}. The higher order corrections to the neutrino Dirac mass are of the form
\begin{equation}
\left(l\nu^c\Psi^2_{\nu}\Psi^{\prime}_{\nu}\right)_{\mathbf{1}}h_u/\Lambda^3\,.
\end{equation}
The corresponding contributions are suppressed by $\lambda^3$ compared to  the LO term $y(l\nu^c)_{\mathbf{1}}h_u$. Such small corrections have a tiny impact for the neutrino mass matrix and lepton mixing parameters, and therefore can be neglected. The NLO corrections to the RH neutrino Majorana mass terms are
\begin{eqnarray}
\nonumber \delta w_{\nu}&=& s_1\left(\nu^c\nu^c\right)_{\mathbf{1}}\left(\chi\chi\right)_{\mathbf{1}}/\Lambda
+s_2\left(\left(\nu^c\nu^c\right)_{\mathbf{2}}\left(\chi\chi\right)_{\mathbf{2}}\right)_{\mathbf{1}}/\Lambda
+s_3\left(\left(\nu^c\nu^c\right)_{\mathbf{3}}\left(\chi\chi\right)_{\mathbf{3}}\right)_{\mathbf{1}}/\Lambda\\
&&+s_4\left(\left(\nu^c\nu^c\right)_{\mathbf{3^{\prime}}}\left(\chi\chi\right)_{\mathbf{3^{\prime}}}\right)_{\mathbf{1}}/\Lambda
+s_5\xi\left(\left(\nu^c\nu^c\right)_{\mathbf{3^{\prime}}}\chi\right)_{\mathbf{1}}/\Lambda
+s_6\xi^2\left(\nu^c\nu^c\right)_{\mathbf{1}}/\Lambda\,.
\end{eqnarray}
The resulting corrections to the RH neutrino mass matrix $m_M$ can be obtained by inserting the LO vacuum of $\chi$ and $\xi$ in Eq.~\eqref{eq:neutrino-vacuum-one} into these operators. Another source of corrections to $m_M$ arises from the LO superpotential $w_{\nu}$ in Eq.~\eqref{eq:neutrino_yukawa_model2} evaluated with the NLO VEVs of Eq.~\eqref{eq:vacuum_NLO_model2}. Adding the two contributions, we obtained the corrected RH neutrino mass matrix as
\begin{equation}
\label{eq:neutrino_mass_NLO}m^{NLO}_M=a\left(\begin{array}{ccc}
    2   &  -1  &  -1  \\
    -1  &  2  &  -1  \\
    -1  &  -1  &  2
\end{array}\right)
+b\left(\begin{array}{ccc}
    0   &  1  &  1  \\
    1  &  1  &  0  \\
    1  &  0  &  1
\end{array}\right)
+c\left(\begin{array}{ccc}
    1   &  0  &  0  \\
    0  &  0  &  1  \\
    0  &  1  &  0
\end{array}\right)
+d\left(\begin{array}{ccc}
    0   &  1  &  -1  \\
    1  &  2  &  0  \\
    -1  &  0  &  -2
\end{array}\right)\,,
\end{equation}
with
\begin{eqnarray}\label{eq:parameter-redefine-one}
\nonumber &&a=y_1v^{\prime}_S+2s_4v^{2}_{\chi}/\Lambda,\quad b=y_2v_{\eta}+s_2v^{2}_{\chi}/\Lambda, \quad c=s_6v^2_{\xi}/\Lambda-2s_1v^2_{\chi}/\Lambda\,, \\
&& d=y_1\delta v_S+s_5v_{\chi}v_{\xi}/\Lambda=\left[y_1\left(\frac{h_4}{g_2}\frac{v_S}{v_{\eta}}-\frac{h_7}{g_2}\right)+s_5\right]
\frac{v_{\chi}v_{\xi}}{\Lambda}\,,
\end{eqnarray}
where parameters $a$ and $b$ have been redefined to include the NLO contributions. Note that $c$ and $d$ arise from the NLO contributions, and they are suppressed by a factor $\lambda$ with respect to $a$ and $b$, i.e.
\begin{equation}
a,b \sim\lambda\Lambda,\qquad c,d\sim\lambda^2\Lambda\,.
\end{equation}
Applying the see-saw relation, the light neutrino mass matrix at NLO takes the form
\begin{eqnarray}
\nonumber m^{NLO}_{\nu}&=&-m_D\left(m^{NLO}_M\right)^{-1}m^T_D\,,\\
&=&\alpha \left(\begin{array}{ccc}
    2   &  -1  &  -1  \\
    -1  &  2  &  -1  \\
    -1  &  -1  &  2
  \end{array}\right)+\beta \left(\begin{array}{ccc}
    1   &  0  &  0  \\
    0  &  0  &  1  \\
    0  &  1  &  0
  \end{array}\right)+\gamma \left(\begin{array}{ccc}
    0   &  1  &  1  \\
    1  &  1  &  0  \\
    1  &  0  &  1
  \end{array}\right)+\delta \left(\begin{array}{ccc}
    0   &  1  &  -1  \\
    1  &  2  &  0  \\
    -1  &  0  &  -2
  \end{array}\right)\,.
\end{eqnarray}
It is the most general neutrino mass matrix invariant under residual family symmetry $G_{\nu}=Z_2^{SU}=\{1,SU\}$, as shown in Eq.~\eqref{eq:general-mass-matrix}. The parameters $\alpha$, $\beta$, $\gamma$ and $\delta$ are given by
\begin{eqnarray}\label{eq:linear1}
\nonumber&&\alpha=\frac{-a(2b+c)+d^2}{(3a-b+c)\left[(3a+b-c)(2b+c)-6d^2\right]}\,,\\
\nonumber&&\beta=\frac{-3a^2-b^2+c^2+2d^2}{(3a-b+c)\left[(3a+b-c)(2b+c)-6d^2\right]}\,,\\
\nonumber&&\gamma=-\frac{3a^2+b(c-b)+d^2}{(3a-b+c)\left[(3a+b-c)(2b+c)-6d^2\right]}\,,\\
&&\delta=-\frac{d}{(3a+b-c)(2b+c)-6d^2}\,,
\end{eqnarray}
where the overall factor $y^2v^2_u$ is omitted here. Because the theory is required to be invariant under the generalised CP transformations, the phases of the model parameters are strongly constrained. The vacuum alignment of Eq.~\eqref{eq:VEVs_relation_model2} implies that the phase different between $v_{\chi}$ and $v_{\xi}$ is $0$, $\pi$ or $\pi/2$ for $g_6g_7>0$ and $g_6g_7<0$ respectively. Further recalling that $v_s$ and $v^2_{\xi}$ should have a common phase (up to relative sign) to avoid degenerate light neutrino masses at LO. Therefore, $a$, $b$ and $c$ are real while $d$ is real or imaginary after the unphysical phase of $v_{\xi}$ is extracted. As a result, $\alpha$, $\beta$ and $\gamma$ in Eq.~\eqref{eq:neutrino_mass_NLO} are real parameters whereas $\delta$ can be real or purely imaginary. In the following, we discuss the two cases one after another.

Firstly, we consider the case that $v_{\chi}$ is real, which corresponds to the parameter domain of $g_6g_7>0$. We can check that the remnant CP symmetry in the neutrino sector is $H^\nu_{CP} =\{\rho _{\mathbf{r}}(1),\rho_{\mathbf{r}}(SU)\}$ in this case. All the four parameters $\alpha$, $\beta$, $\gamma$ and $\delta$ are real. This is exactly the case (I) of model-independent analysis in section~\ref{sec:model_independent}.
Remembering that the subleading operators in the charged lepton sector induce corrections to the lepton mixing angles as small as $\lambda^3$. Hence, the lepton flavor mixing is determined by the neutrino sector. From section~\ref{sec:model_independent}, we know that the resulting lepton mixing matrix is
\begin{eqnarray}
\label{eq:UPMNS3}U_{PMNS}=\left(
\begin{array}{ccc}
 \sqrt{\frac{2}{3}} & \frac{\cos\theta }{\sqrt{3}} & \frac{\sin\theta }{\sqrt{3}} \\
 -\frac{1}{\sqrt{6}} & \frac{\cos\theta }{\sqrt{3}}+\frac{\sin\theta }{\sqrt{2}} & -\frac{\cos\theta }{\sqrt{2}}+\frac{\sin\theta }{\sqrt{3}} \\
 -\frac{1}{\sqrt{6}} & \frac{\cos\theta }{\sqrt{3}}-\frac{\sin\theta }{\sqrt{2}} & \frac{\cos\theta }{\sqrt{2}}+\frac{\sin\theta }{\sqrt{3}}
\end{array}
\right)\,,
\end{eqnarray}
with
\begin{eqnarray}
\tan2\theta =\frac{-2\sqrt{6}\delta}{3\alpha-2\beta-\gamma}=\frac{2\sqrt{6}d}{3a-b-2c}\sim\mathcal{O}(\lambda)\,.
\end{eqnarray}
The lepton mixing angles are given by
\begin{eqnarray}
\nonumber&\sin\theta_{13}=\left|\frac{\sin\theta}{\sqrt{3}}\right|\simeq\left|\frac{\sqrt{2}d}{3a-b-2c}\right|\sim\mathcal{O}(\lambda)\,,\\
&\sin^2\theta_{12}\simeq\frac{1}{3}+\mathcal{O}(\lambda^2),\qquad \sin^2\theta_{23}\simeq\frac{1}{2}\pm\frac{2d}{3a-b-2c}\,.
\end{eqnarray}
We see that the reactor angle $\theta _{13}$ is predicted to be of the correct order of $\lambda$, and thus experimentally preferred value could be achieved. The solar mixing angle retains its tri-bimaximal value to the first order of $\lambda$, and the atmospheric angle can deviate from its  maximal mixing value of $45^{\circ}$. As a consequence, the deviation of the atmospheric angle from maximal mixing, indicated by the latest global fits, can be produced. In addition, we find a simple sum rule $\sin^2\theta_{23}\simeq0.5\pm\sqrt{2}\;\sin\theta_{13}$. This relation might be testable in the near future as soon as the experimental uncertainties for $\theta_{23}$ are reduced. Furthermore, since the light neutrino mass matrix is real, there is no CP violation in this case, both the Dirac CP phase and the Majorana CP phases are $0$ or $\pi$.

Then we consider the remaining case of $v_{\chi} $ being purely imaginary, i.e. the phase different between $v_{\chi}$ and $v_{\xi}$ is $\pm\frac{\pi}{2}$. This scenario can be realized in the parameter domain $g_6g_7<0$. The generalised CP symmetry is broken down to $H^{\nu}_{CP} =\{\rho_{\mathbf{r}}(S), \rho_{\mathbf{r}}(U)\}$ in the neutrino sector. This corresponds to the case (II) of section~\ref{sec:model_independent}. The resulting parameters $\alpha$, $\beta$, $\gamma$ are real and $\delta$ is imaginary. The lepton mixing matrix is of the form
\begin{eqnarray}\label{eq:UPMNS4}
U_{PMNS}=\left(
\begin{array}{ccc}
 \sqrt{\frac{2}{3}} & \frac{\cos\theta }{\sqrt{3}} & \frac{\sin\theta }{\sqrt{3}} \\
 -\frac{1}{\sqrt{6}} & \frac{\cos\theta }{\sqrt{3}}+\frac{i \sin\theta }{\sqrt{2}} & -\frac{i \cos\theta }{\sqrt{2}}+\frac{\sin\theta }{\sqrt{3}} \\
 -\frac{1}{\sqrt{6}} & \frac{\cos\theta }{\sqrt{3}}-\frac{i \sin\theta }{\sqrt{2}} & \frac{i \cos\theta }{\sqrt{2}}+\frac{\sin\theta }{\sqrt{3}}
\end{array}
\right)\,,
\end{eqnarray}
with
\begin{eqnarray}
\tan2\theta=\frac{2i\sqrt{6}\delta}{3\left(\alpha+\gamma\right)}= \frac{2i\sqrt{6}d}{3\left(a+b\right)}\sim\mathcal{O}\left(\lambda\right)\,.
\end{eqnarray}
Consequently the three mixing angles $\theta_{13}$, $\theta_{12}$ and $\theta_{23}$ are modified to
\begin{eqnarray}
\sin\theta_{13}\simeq\left|\frac{\sqrt{2}d}{3\left(a+b\right)}\right|\sim\mathcal{O}(\lambda), \quad \sin^2\theta_{12}=\frac{1}{3}+\mathcal{O}(\lambda^2),\quad \sin^2\theta_{23}=\frac{1}{2}\,.
\end{eqnarray}
It is noteworthy that we obtain maximal Dirac CP violation $\delta_{CP}=\pm\pi/2$ in this case while the Majorana CP phases are still trivial with $\sin\alpha_{21}=\sin\alpha_{13}=0$. In short summary, our model produces the tri-bimaximal mixing at LO, which is further broken down to trimaximal $\text{TM}_1$ mixing by NLO contributions. Depending on the
coupling product $g_6g_7$ being positive or negative, the two cases arising from the model independent analysis can be realized.

\section{\label{sec:conclusion}Conclusions}

The measurement of sizable reactor mixing angle $\theta_{13}$ has opened up the possibility of measuring leptonic CP violations. In particular, the measurement of Dirac CP phase is one of the primary goals of next
generation neutrino oscillation experiments. On the theoretical side, the origin of CP violation remains a mystery. Extending family symmetry to include generalised CP symmetry together with its spontaneous breaking is a promising framework to predict both mixing angles and CP phases.

In this work, we analyse the interplay of generalised CP symmetry and the $S_4$ family symmetry. Firstly we perform a model independent analysis of the possible lepton mixing matrices and the corresponding lepton mixing parameters, which arise from the symmetry breaking of $S_4\rtimes H_{CP}$ into $Z^{T}_3\rtimes H^{l}_{CP}$ in the charged lepton sector and $Z^{SU}_2\rtimes H^{\nu}_{CP}$ in the neutrino sector. We find that the lepton flavor mixing is of the $\text{TM}_1$ form and the Dirac CP can be vanishing or maximally broken while the Majorana CP is trivial with $\sin\alpha_{21}=\sin\alpha_{31}=0$.

Furthermore, we construct two models to realize the above model independent results based on $S_4$ family symmetry and the generalised CP symmetry. The two models differ in the neutrino sectors. In the first model, the flavon fields enter in the neutrino Dirac mass term instead of the Majorana mass term for right-handed neutrinos at LO. The resulting light neutrino mass matrix is predicted to depend on three real parameters, and therefore the absolute neutrino masses and the effective mass $\left|m_{\beta\beta}\right|$ for neutrinoless double beta decay are completely fixed after considering the constraints from the measured values of the neutrino mass squared differences $\Delta m^2_{sol}$ and $\Delta m^2_{atm}$ and the reactor angle $\theta_{13}$. The lepton mixing matrix is the $\text{TM}_1$ pattern, and the subleading corrections are small enough to be negligible. In the case of $g_3g_4\left(g_3g_5+g_2g_6\right)>0$, the Dirac CP phase $\delta_{CP}$ is $0$ or $\pi$, and neutrino mass spectrum can be normal ordering or inverted ordering. For the case of $g_3g_4\left(g_3g_5+g_2g_6\right)<0$, the Dirac CP is maximal $\delta_{CP}=\pm\pi/2$, and the neutrino mass spectrum can only be normal ordering.

In the second model, the $S_4$ family symmetry is broken down to $Z^{S}_2\times Z^{SU}_2$ in the neutrino sector at LO, and therefore the LO lepton mixing is of the tri-bimaximal form. NLO correction terms break the remnant symmetry $Z^{S}_2\times Z^{SU}_2$ into $Z^{SU}_2$, as a result, the $\text{TM}_1$ mixing is produced and the relative smallness of $\theta_{13}$ with respect to $\theta_{12}$ and $\theta_{23}$ is explained. Depending on the product $g_6g_7$ being positive or negative, the Dirac CP is predicted to be conserved or maximally broken. Moreover, we have shown that the desired vacuum alignment together with their phase structure can be achieved.

\section*{Acknowledgements}

One of the author (G.J.D.) is grateful to Stephen F. King, Christoph Luhn and Alexander J. Stuart for for stimulating discussions on generalised CP symmetry. G.J.D. would also like to thank Stephen F. King and the School of Physics and Astronomy at the University of Southampton for hospitality during his visit, where part of this work was done. The research was partially supported by  the National Natural Science Foundation of China under Grant Nos. 11275188 and 11179007.

\newpage

\section*{Appendix}
\cleqn

\begin{appendix}

\section{\label{sec:A}Group Theory of $S_4$}

\begin{table}[t!]
\begin{center}
\begin{tabular}{|c|c|c|c|}\hline\hline
 ~~  &  $S$  &   $T$    &  $U$  \\ \hline
~~~${\bf 1}$, ${\bf 1^\prime}$ ~~~ & 1   &  1  & $\pm1$  \\ \hline
   &   &    &    \\ [-0.16in]
${\bf 2}$ &  $\left( \begin{array}{cc}
    1&0 \\
    0&1
    \end{array} \right) $
    & $\left( \begin{array}{cc}
    \omega&0 \\
    0&\omega^2
    \end{array} \right) $
    & $\left( \begin{array}{cc}
    0&1 \\
    1&0
    \end{array} \right)$\\ [0.12in]\hline
   &   &    &    \\ [-0.16in]
${\bf 3}$, ${\bf 3^\prime}$ & $\frac{1}{3} \left(\begin{array}{ccc}
    -1& 2  & 2  \\
    2  & -1  & 2 \\
    2 & 2 & -1
    \end{array}\right)$
    & $\left( \begin{array}{ccc}
    1 & 0 & 0 \\
    0 & \omega^{2} & 0 \\
    0 & 0 & \omega
    \end{array}\right) $
    & $\mp\left( \begin{array}{ccc}
    1 & 0 & 0 \\
    0 & 0 & 1 \\
    0 & 1 & 0
    \end{array}\right)$
\\[0.22in] \hline\hline
\end{tabular}
\caption{\label{tab:representation}The representation matrices of the generators $S$, $T$ and $U$ for the five irreducible representations of $S_4$ in the chosen basis, where $\omega=e^{2\pi i/3}$. }
\end{center}
\end{table}

$S_4$ is the permutation group of order 4 with 24 elements, and it has been widely used as a family symmetry. In this work, we shall follow the conventions and notations of Refs.~\cite{Ding:2013hpa,King:2010}, where $S_4$ is expressed in terms of three generators $S$, $T$ and $U$. These three generators satisfy the multiplication rules:
\begin{eqnarray}
S^2=T^3=U^2=(ST)^3=(SU)^2=(TU)^2=(STU)^4=1\,.
\end{eqnarray}
Note that the generators $S$ and $T$ alone generate the group $A_4$, while the generated group by $T$ and $U$ is $S_3$. The $S_4$ group elements can be divided into 5 conjugacy classes
\begin{eqnarray}
\nonumber 1C_1 &=& \left\{1 \right\}, \\
\nonumber 3C_2 &=& \left\{S, TST^2, T^2ST \right\},  \\
\nonumber 6C_2^\prime &=& \left\{U,TU,SU,T^2U,STSU,ST^2SU\right\},  \\
\nonumber 8C_3 &=& \left\{T,ST,TS,STS,T^2,ST^2,T^2S,ST^2S\right\},  \\
6C_4 &=& \left\{STU,TSU,T^2SU,ST^2U,TST^2U,T^2STU\right\}\,,
\end{eqnarray}
where the conjugacy class is denoted by $kC_n$, $k$ is the number of elements belonging to it, and the subscript $n$ is the order of the elements contained in it.  As a result of these conjugacy classes and the theorems that prove that the number of inequivalent irreducible representations is equal to the number of conjugacy classes and the sum of the squares of the dimensions of the inequivalent irreducible representations must be equal to the order of the group,  it is easy to see that $S_4$ has two singlet irreducible representations $\mathbf{1}$ and $\mathbf{1^{\prime}}$, one two-dimensional representation $\mathbf{2}$ and two three-dimensional irreducible representations $\mathbf{3}$ and $\mathbf{3^{\prime}}$. In this work, we shall work in the basis where the representation matrix of the generator $T$ is diagonal. As a result, the charged lepton mass matrix would be diagonal if the remnant subgroup $Z^{T}_3\equiv\left\{1, T, T^2\right\}$ is preserved in the charged lepton sector. The explicit forms of the representation matrix for the three generators are listed in Table~\ref{tab:representation}, and hence the chosen basis coincides with that of Ref.~\cite{Ding:2013hpa}. The character table of $S_4$ group follows immediately, as shown in Table~\ref{tab:characher}. Moreover, the Kronecker products between different irreducible representations are as follows
\begin{eqnarray}
\nonumber && \bf{1}\otimes \mathbf{R}=\mathbf{R},~~\bf{1^\prime}\otimes \bf{1^\prime}=\bf{1},~~ \bf{1^\prime}\otimes\bf{2}=\bf{2},~~ \bf{1^\prime}\otimes\bf{3}=\bf{3^\prime},~~ \bf{1^\prime}\otimes\bf{3^\prime}=\bf{3},  \\
\nonumber && \bf{2}\otimes\bf{2}=\bf{1}\oplus\bf{1^\prime}\oplus\bf{2},~~\bf{2}\otimes\bf{3}=\bf{2}\otimes\bf{3^\prime}=\bf{3}\otimes\bf{3^\prime},\\
&& \bf{3}\otimes\bf{3}=\bf{3^\prime}\otimes\bf{3^\prime}=\bf{1}\oplus\bf{2}\oplus\bf{3}\oplus\bf{3^\prime},~~ \bf{3}\otimes\bf{3^\prime}=\bf{1^\prime}\oplus\bf{2}\oplus\bf{3}\oplus\bf{3^\prime}
\end{eqnarray}
where $\mathbf{R}$ stands for any irreducible representation of $S_4$.

\begin{table}[t!]
\begin{center}
\begin{tabular}{|c|c|c|c|c|c|} \hline \hline
  Classes & $1C_1$  & $3C_2$  & $6C_2^\prime$  & $8C_3$  &  $6C_4$ \\ \hline
  $G$  &  1  &  $S$  &  $U$  & $T$  &   $STU$   \\ \hline
  $\bf{1}$  &  1  &  1  &  1  &  1  &  1  \\
  $\bf{1^\prime}$  &  1  &  1  &  $-1$  &  1  &  $-1$  \\
  $\bf{2}$  &  2  &  2  &  0  &  $-1$  &  0  \\
  $\bf{3}$  &  3  &  $-1$  &  $-1$  &  0  &  1  \\
  $\bf{3^\prime}$  &  3  &  $-1$  &  1  &  0  &  $-1$  \\ \hline \hline
\end{tabular}
\caption{\label{tab:characher}Character table of the group $S_4$, where $G$ denotes the representative element of each conjugacy class.}
\end{center}
\end{table}

In the end, we present the Clebsch-Gordan (CG) coefficients in the chosen basis. All the CG coefficients can be reported in the form of $\alpha\otimes \beta$, $\alpha_i$ denotes the element of the left base vectors $\alpha$, and $\beta_i$ is the element of the right base vectors $\beta$. For the product of the singlet $\mathbf{1^{\prime}}$ with a doublet or a triplet, we have
\begin{equation}
\begin{array}{lll}
\bf{1^\prime}\otimes\bf{2}=\bf{2}=\alpha\left(\begin{array}{c}
                                          \beta_1 \\
                                          -\beta_2
                                          \end{array}\right),\quad
\bf{1^\prime}\otimes\bf{3}=\bf{3^\prime} =\alpha\left(\begin{array}{c}
                                                  \beta_1  \\
                                                  \beta_2  \\
                                                  \beta_3
                                                  \end{array}\right),\quad
\bf{1^\prime}\otimes\bf{3^\prime}=\bf{3}=\alpha\left(\begin{array}{c}
                                                  \beta_1  \\
                                                  \beta_2  \\
                                                  \beta_3
                                                  \end{array}\right)
\end{array}\,.
\end{equation}
The CG coefficients for the products involving the doublet   representation $\mathbf{2}$ are found to be
\begin{eqnarray*}
\begin{array}{lll}
\bf{2}\otimes\bf{2}=\bf{1}\oplus\bf{1^\prime}\oplus\bf{2},& \qquad\qquad &
\text{with}\qquad \left\{\begin{array}{l}
                  \bf{1}\;=\alpha_1\beta_2+\alpha_2\beta_1  \\
                  \bf{1^\prime}=\alpha_1\beta_2-\alpha_2\beta_1  \\
                  \bf{2}\;=\left(\begin{array}{c}
                         \alpha_2\beta_2  \\
                         \alpha_1\beta_1
                         \end{array}\right)
                  \end{array}\right. \\
                  \\[-8pt]
\bf{2}\otimes\bf{3}=\bf{3}\oplus\bf{3^\prime},& \qquad\qquad &
      \text{with}\qquad \left\{\begin{array}{l}
                  \bf{3}\;=\left(\begin{array}{c}
                          \alpha_1\beta_2+\alpha_2\beta_3 \\
                          \alpha_1\beta_3+\alpha_2\beta_1  \\
                          \alpha_1\beta_1+\alpha_2\beta_2
                          \end{array}\right) \\[0.25in]
                  \bf{3^\prime}=\left(\begin{array}{c}
                          \alpha_1\beta_2-\alpha_2\beta_3 \\
                          \alpha_1\beta_3-\alpha_2\beta_1  \\
                          \alpha_1\beta_1-\alpha_2\beta_2
                          \end{array}\right) \\
                 \end{array}\right.\\
                 \\[-8pt]
\bf{2}\otimes\bf{3^\prime}=\bf{3}\oplus\bf{3^\prime},& \qquad\qquad &
      \text{with}\qquad\left\{\begin{array}{l}
                  \bf{3}\;=\left(\begin{array}{c}
                          \alpha_1\beta_2-\alpha_2\beta_3 \\
                          \alpha_1\beta_3-\alpha_2\beta_1  \\
                          \alpha_1\beta_1-\alpha_2\beta_2
                          \end{array}\right) \\[0.25in]
                  \bf{3^\prime}=\left(\begin{array}{c}
                          \alpha_1\beta_2+\alpha_2\beta_3 \\
                          \alpha_1\beta_3+\alpha_2\beta_1  \\
                          \alpha_1\beta_1+\alpha_2\beta_2
                          \end{array}\right) \\
\end{array}\right.\\
\end{array}
\end{eqnarray*}
Finally, for the products of the triplet representations $\mathbf{3}$ and $\mathbf{3^{\prime}}$, we find
\begin{eqnarray*}
\begin{array}{ll}
     \bf{3}\otimes\bf{3}=\bf{3^\prime}\otimes\bf{3^\prime}=\bf{1}\oplus\bf{2}\oplus\bf{3}\oplus\bf{3^\prime},&\quad
      \text{with}\quad\left\{\begin{array}{l}
                 \bf{1}\;=\alpha_1\beta_1+\alpha_2\beta_3+\alpha_3\beta_2  \\[0.1in]
                 \bf{2}\;=\left(\begin{array}{c}
                         \alpha_2\beta_2+\alpha_1\beta_3+\alpha_3\beta_1  \\
                         \alpha_3\beta_3+\alpha_1\beta_2+\alpha_2\beta_1
                        \end{array}\right)\\[0.18in]
                 \bf{3}\;=\left(\begin{array}{c}
                        \alpha_2\beta_3-\alpha_3\beta_2  \\
                        \alpha_1\beta_2-\alpha_2\beta_1  \\
                        \alpha_3\beta_1-\alpha_1\beta_3
                        \end{array}\right) \\[0.25in]
                 \bf{3^\prime}=\left(\begin{array}{c}
                        2\alpha_1\beta_1-\alpha_2\beta_3-\alpha_3\beta_2  \\
                        2\alpha_3\beta_3-\alpha_1\beta_2-\alpha_2\beta_1  \\
                        2\alpha_2\beta_2-\alpha_3\beta_1-\alpha_1\beta_3
                        \end{array}\right) \\
                 \end{array}\right.\\
                 \\[-8pt]
\bf{3}\otimes\bf{3^\prime}=\bf{1^\prime}\oplus\bf{2}\oplus\bf{3}\oplus\bf{3^\prime},&\quad
      \text{with}\quad\left\{\begin{array}{l}
                 \bf{1}\;=\alpha_1\beta_1+\alpha_2\beta_3+\alpha_3\beta_2  \\[0.1in]
                 \bf{2}\;=\left(\begin{array}{c}
                         \alpha_2\beta_2+\alpha_1\beta_3+\alpha_3\beta_1  \\
                         -(\alpha_3\beta_3+\alpha_1\beta_2+\alpha_2\beta_1)
                        \end{array}\right)\\[0.18in]
                 \bf{3}\;=\left(\begin{array}{c}
                        2\alpha_1\beta_1- \alpha_2\beta_3-\alpha_3\beta_2  \\
                        2\alpha_3\beta_3- \alpha_1\beta_2-\alpha_2\beta_1  \\
                        2\alpha_2\beta_2-\alpha_3\beta_1-\alpha_1\beta_3
                        \end{array}\right) \\[0.25in]
                 \bf{3^\prime}=\left(\begin{array}{c}
                       \alpha_2\beta_3-\alpha_3\beta_2  \\
                        \alpha_1\beta_2-\alpha_2\beta_1  \\
                        \alpha_3\beta_1-\alpha_1\beta_3
                        \end{array}\right) \\
                 \end{array}\right.\\
     \end{array}
\end{eqnarray*}
We note that the CG coefficients presented above are in accordance with the results of Refs.~\cite{Ding:2013hpa,King:2010,Ding:2009iy}.

\end{appendix}

\end{document}